\documentclass[preprint,pre,amsmath]{revtex4-2}
\usepackage[pdftex]{graphicx} 
\usepackage{color}
\usepackage{bm}
\usepackage{epsfig}
\usepackage{latexsym}
\usepackage{nameref,hyperref}
\usepackage{cases,setspace,amsmath,soul}
\usepackage{wasysym}

\usepackage{amsmath,hyperref}

\def\be{\begin{equation}}
\def\ee{\end{equation}}
\def\bea{\begin{eqnarray}}
\def\eea{\end{eqnarray}}
\def\bi{\begin{itemize}}
\def\ei{\end{itemize}}

\def\bsn{\begin{subnumcases}}
\def\esn{\end{subnumcases}}
\def\bml{\begin{mathletters}}
\def\eml{\end{mathletters}}

\def\nn{\nonumber}

\begin{document}

\title[Run-and-tumble]{Run-and-tumble particle with saturating rates}
\author{Kavita Jain(1,*) and Sakuntala Chatterjee(2,*)} 
\email{(1) jain@jncasr.ac.in, (2) sakuntala.chatterjee@bose.res.in}
\affiliation{(1) Theoretical Sciences Unit, Jawaharlal Nehru Centre for Advanced Scientific Research, Bangalore 560064, India \\
(2) Department of Physics of Complex Systems, S.N. Bose National Center for Basic Sciences, Salt Lake Sector 3, Kolkata 700106, India \\
* equal contribution}

\clearpage

\begin{abstract}
We consider a run-and-tumble particle whose speed and tumbling rate are space-dependent on an infinite line. Unlike most of the previous work on such models, here we make the physical assumption that at large distances, these rates saturate to a constant. For our choice of rate functions, we show that a stationary state exists, and the exact steady state distribution decays exponentially or faster and can be unimodal or bimodal. The effect of boundedness of rates is seen in the mean-squared displacement of the particle that displays qualitative features different from those observed in the previous studies where it approaches the stationary state value monotonically in time; in contrast, here we find that if the initial position of the particle is sufficiently far from the origin, the variance in its position either varies nonmonotonically or plateaus before reaching the stationary state. These results are captured quantitatively by the exact solution of the Green's function when the particle has uniform speed but the tumbling rates change as a step-function in space; the insights provided by this limiting case are found to be consistent with the numerical results for the general model. 
\end{abstract}

\maketitle

\clearpage


\section{Introduction}

Active matter comprises a wide class of natural systems ranging from flock of birds to bacterial colony as well as artificially prepared systems like engineered {\sl E.coli} bacteria \cite{arlt2019dynamics, arlt2018painting, frangipane2018dynamic} or artificial Janus colloids  \cite{bregulla2014stochastic, jahanshahi2020realization, soker2021activity, auschra2021density}, and have received a lot of attention lately \cite{schweitzer2003brownian, romanczuk2012active, bechinger2016active, ramaswamy2017active}. In such systems, each particle consumes  energy from its environment and performs an activity or active motion which is not symmetric under time-reversal \cite{Fodor:2015,Fodor:2016,Smith:2023} and hence active systems are intrinsically out-of-equilibrium even at the level of a single particle.

One of the most well-studied active motion is run-and-tumble motion \cite{Weiss:2002, Masoliver:2017, Novikova:2017, Doering:2018} which is seen among various different kinds of microorganisms including prokaryotic cells like {\sl E. coli, S. typhimurium, B. subtilis, R. sphaeroides and S. marcescens} \cite{berg2004coli, galloway1980histidine, sidortsov2017role, rosser2014modelling, ariel2015swarming} or eukaryotic cells such as {\sl C. rheinhartii or T. foetus} \cite{polin2009chlamydomonas, lenaghan2014unlocking}. During a run, a directed motion takes place while during a tumble, the direction of motion is randomized, and by repeatedly switching between run mode and tumble mode, an organism can navigate in its environment. Experiments have shown that the run durations are exponentially distributed \cite{block1982impulse} which means that the noise present in run-and-tumble motion due to abrupt changes in the run direction is not the Gaussian white noise, but is exponentially correlated in time.

A run-and-tumble motion is defined by specifying the velocity or run speed and the tumbling rate. These rates can be intrinsic to the active particle or they can be influenced by the environment; for example, during bacterial chemotaxis, {\sl E.coli} bacteria uses run-and-tumble motility to migrate towards regions with higher nutrient concentration \cite{berg2004coli}. Experiments suggest that while the run speed of the cell generally remains uniform \cite{Berg:1972, macnab1972gradient}, the tumbling rate depends on the local concentration gradient of the nutrient in such a way that uphill runs are extended and downhill runs are shortened \cite{Chatterjee:2011, tu2013quantitative, dev2018optimal, Schnitzer:1993}, and therefore, by changing the nutrient environment, tumbling rates can be modified. However, even in the absence of external stimulus, certain microorganisms show a change in their tumbling rate \cite{chou2003interplay}. For examples, in bacterial cells like {\sl Myxococcus xanthus} or eukaryotic cells like {\sl Dictyostelium discoideum} \cite{kearns1998mx, shi1996cell}, each cell secretes a chemical and the tumbling rate depends on this chemical trail along the cell trajectory. These cells use chemotaxis as well as chemokinesis to move around: while the tumbling rate is sensitive to the chemical concentration gradient during chemotaxis, the absolute value of the chemical concentration at the cell position regulates the tumbling rate during chemokinesis. Recently, it was found that even the run speed can be changed; some examples include an engineered strain of {\sl E.coli} bacteria which responds to the intensity of light by modulating its run speed \cite{arlt2019dynamics, arlt2018painting, frangipane2018dynamic, caprini2022dynamics, vizsnyiczai2017light, stenhammar2016light, caprini2020activity} and wild-type bacteria that changes its speed on sensing the ligand  \cite{Naaz:2021}.

The above mentioned experiments suggest that, in general, the active particle responds to the changes in its environment by modifying both the run speed and tumbling rate. Furthermore, it is expected that a strong signal from the environment will elicit a strong response from the particle; for example, a steep chemical gradient will bring about a large change in the tumbling rate, or by shining a high intensity light, the particle velocity will also show a larger increase. However, neither the tumbling rate nor the velocity can increase in an unbounded manner with the strength of the input signal and beyond a limit, the response of any physical system is expected to saturate. Yet most models assume that either the speed or tumbling rate (and not both) are inhomogeneous and, importantly, these rates increase without bound in space  \cite{Doering:2018,Dor:2019,Angelani:2019,Singh:2020,Sandev:2022}.

Motivated by the above physical and biological considerations, in this work, we consider a simple model of a run-and-tumble particle in one dimension with position-dependent velocity and tumbling rate which saturate at large distances from the origin. As we are mainly interested in understanding the effect of spatial variation in rates on the stationary state and how the bounded nature of rates affect the dynamics of the active particle, our choice of rates is  made keeping these general questions in mind rather than modeling a specific biological system or situation. As described in Sec.~\ref{ch2}, the particle moves with rightward velocity $v_+(x)$ and leftward velocity $v_-(x)$, while the corresponding tumbling rates are $\gamma_+(x)$ and $\gamma_-(x)$, respectively. The rates $v_\pm(x)$ and $\gamma_\pm(x)$ given by (\ref{vchoice}) and (\ref{gchoice}), respectively, saturate with $x$ for large $|x|$ to zero or a nonzero constant.  In Sec.~\ref{statstate}, we first identify the parameter region where the stationary state exists, and find the exact position distribution which decays exponentially or faster, and it may be unimodal centered around the origin, or bimodal with two peaks symmetrically placed on both sides of the origin. 

In Sec.~\ref{sec:dyn}, to understand the dynamics of the active particle far from its steady state, we focus on the mean-squared displacement (MSD) of the particle as a function of time, starting from a given initial position. Our Monte Carlo simulations show that if the particle starts close to the origin, the MSD increases monotonically towards the stationary state value. But on starting sufficiently far from the origin, the dynamics of the MSD fall in three distinct dynamical regimes: at short times, it grows with time as a power law with an exponent $2$ or $3$, depending on the initial orientation of the particle; at intermediate times, it reaches a plateau or a peak, depending on whether or not the tumbling rates vanish at large distances; and at large times, the MSD  relaxes exponentially fast to its steady state value. To the best of our knowledge, a plateau or peak in the MSD at intermediate times has not been reported earlier for a run-and-tumble motion, and is one of the most striking effects of saturating rates.  We obtain an analytical understanding of these results in a limiting case where the speeds are homogeneous in space and the tumbling rates vary as a step-function at the origin by calculating an exact expression of the Green's function.  

\section{Description of the model}
\label{ch2}

We consider a run-and-tumble particle (RTP) in one dimension which is described by its position $-\infty < x < \infty$ and  internal states, {\it viz.}, $+$ (right-mover) and $-$ (left-mover) in continuous time. In time $dt \to 0$, the right-mover at position $x$ at time $t$ either moves rightward at rate $v_+(x)$ or changes its internal state to $-$ at rate $\gamma_+(x)$; similarly, the left-mover  either moves leftward or reverses its direction at rate $v_-(x)$ and $\gamma_-(x)$, respectively. 
Then the probability distribution $P_+(x,t)$ and $P_-(x,t)$, respectively, for the positions of the right- and left-mover evolve according to \cite{Weiss:2002}
\bea
\frac{\partial P_+(x,t)}{\partial t} &=& -[v_+(x) P_+(x,t)]'+ \gamma_-(x) P_-(x,t) -\gamma_+(x) P_+(x,t) \label{Pp}\\
\frac{\partial P_-(x,t)}{\partial t} &=& -[-v_-(x) P_-(x,t)]'+ \gamma_+(x) P_+(x,t) -\gamma_-(x) P_-(x,t) \label{Pm}
\eea
where prime denotes the derivative with respect to $x$. The probability distribution of the particle's position is then given by  $P(x,t)=P_+(x,t)+P_-(x,t)$ with $\int_{-\infty}^\infty dx P(x,t)=1$ at all times.

We now assume the following speed and tumble rates that are space-dependent, bounded and asymmetric in the two directions, 
\bea
v_\pm(x) &=& v_0 \mp v_1 \tanh(u x)  \label{vchoice} \\
\gamma_\pm(x) &=&\gamma_0 \pm  \gamma_1\tanh(g x) \label{gchoice}
\eea
where $v_0 \geq v_1 > 0, \gamma_0 \geq \gamma_1 > 0$ as the rates must be non-negative. In the above rates, there are six free parameters, and although we can reduce them to four by scaling time and space, we do not do so as it is helpful to see their explicit dependence. In general, the parameters $u$ and $g$ can take either sign, but here we assume that $u \geq 0, g \geq 0$. From (\ref{vchoice}) and  (\ref{gchoice}),  it follows that the particle moves with higher speed and/or tumbles less often as it approaches the origin, and when it moves away from the origin, it moves slower and/or tumbles more often. 

For $u=g=0$, we obtain the RTP model with homogeneous rates which has been extensively studied \cite{Weiss:2002}. For the parameters $u \neq 0, g=0$, we can write down the Langevin equation of the RTP in the overdamped limit as $\dfrac{dx}{dt} =f(x) + v_0 \eta (t) $, where we have set the damping coefficient to unity; here, the external force $f(x)=v_1 \tanh(u x)$ while $ \eta (t) $ represents the telegraphic noise which takes values $\pm 1$ and flips its sign at a constant rate $\gamma_0$. The external potential then has the form 
\be
\begin{aligned}
V(x) &= -\int_0^x dx' f(x')=\frac{v_1}{u} \ln(\cosh(ux))\label{potential}  
&\sim
\begin{cases}
\frac{v_1 u}{2} x^2 ~&,~ u |x| \ll 1 \\
v_1 |x| ~&,~ u |x| \gg 1
\end{cases}
\end{aligned}
\ee
For $u \to \infty$ where the external potential changes discontinuously at the origin, the stationary state has been investigated \cite{Sevilla:2019,Dhar:2019}. For $u=0, g \neq 0$, the stationary state has  been obtained \cite{Monthus:2021,Bressloff:2021}, and some aspects of the dynamics for $g \to \infty$ have been studied \cite{Singh:2020}. 

Furthermore, unlike in most previous work where the RTP is initially at the origin and the initial density of the left- and right-mover is symmetric, here we assume that 
\be
P_\pm(x,0)=a_\pm \delta(x-x_0) \label{initialcond}
\ee
where $a_+ + a_-=1, 0 \leq a_\pm \leq 1$. Without loss of generality, below we also assume that the initial location $x_0 \geq 0$.

We studied the model described above analytically using (\ref{Pp}) and (\ref{Pm}), and also conducted extensive Monte Carlo simulations in which during a time interval $dt$, depending on its internal state, the particle at position $x$ either switches its direction with probability $\gamma_\pm (x)dt$ or persists in the same direction with velocity $v_\pm(x)$ and moves to the new position $x \pm v_\pm (x)dt$. For the initial condition (\ref{initialcond}), multiple stochastic trajectories $\{x(t)\}$ for the particle's position were generated, and  its mean $\overline{x}(t)$ and MSD, $ \sigma^2(x_0, t)={(\overline{x(t)-x_0)^2}} - (\overline{x}(t)-x_0)^2$ (where the bar denotes an average over independent particle trajectories) were computed. The stationary distribution [see (\ref{ssexact}) below] was also verified for representative set of parameters in simulations (data not shown). In the following sections, we first describe the stationary state and then turn to the dynamics. 

\section{Stationary state of the model}
\label{statstate}


\subsection{Stationary state distribution}
\label{ssp}

For the rates in  (\ref{vchoice}) and (\ref{gchoice}), we first find the regions in the $u-g$ plane where the stationary state is expected to exist.  When $u=g=0$, all the rates are constant in space and we have the telegrapher's equation for which there is no stationary state \cite{Weiss:2002}. But when $u > 0, g=0$,  the stationary state may exist as the speed of the right (left)-mover decreases  with increasing (decreasing) $x$ which prevents the particle to drift away to $x \to \infty (-\infty)$. If now the tumbling rate is also nonuniform and $g > 0$, the right (left)-mover reverses its direction at a higher rate with increasing (decreasing) $x$ which further supports the confinement of the particle. 
On the other hand, for $u < 0, g < 0$, as the right (left)-mover travels faster and reverses its direction less often with increasing (decreasing) $x$, the particle can escape and the stationary state does not exist. 
For $u > 0, g < 0$, while the speed of the right (left)-mover decreases with increasing $x$, it also 
tumbles less often, and the steady state is expected if the particle's speed slows down fast enough to counter the persistent motion; a similar argument applies to $u < 0, g > 0$ regime.

In this work, we consider non-negative $u$ and $g$ only, and show below that a stationary state exists for all $u, g \geq 0$ except when $u=g=0$. As described in Appendix~\ref{app_ss},  the stationary state distribution of the particle's position for the model described in Sec.~\ref{ch2} is given by 
\be
P(x)=  \frac{1}{{\cal N}} \left[\frac{1}{v_+(x)} +\frac{1}{v_-(x)} \right]e^{\int_0^x \left( \frac{\gamma_-(y)}{v_-(y)}- \frac{\gamma_+(y)}{v_+(y)} \right) dy} \label{ssexact}
\ee
provided the speeds $v_\pm(x)$ are nonzero everywhere and the normalization constant 
\be
{\cal N}=\int_{-\infty}^\infty dx  \left[\frac{1}{v_+(x)} +\frac{1}{v_-(x)} \right]e^{\int_0^x \left( \frac{\gamma_-(y)}{v_-(y)}- \frac{\gamma_+(y)}{v_+(y)} \right) dy} 
\label{norm}
\ee
is finite \cite{Schnitzer:1993,Monthus:2021}. Therefore, except for $u \to \infty, v_0=v_1$ for which the speeds can vanish, we first write ${\cal N}=\int_{-\infty}^\infty dx H(x) e^{I(x)}$ where
 \bea
H(x) &=& \frac{1}{v_+(x)} +\frac{1}{v_-(x)}=\frac{2 v_0}{v_0^2-v_1^2 \tanh^2(u x)} \label{Hdefn} \\
I(x) &=& \int_0^x \left( \frac{\gamma_-(y)}{v_-(y)}- \frac{\gamma_+(y)}{v_+(y)} \right) dy=-2 \int_0^x dy \frac{\gamma_0 v_1 \tanh(u y)+ \gamma_1 v_0\tanh(g y)}{v_0^2-v_1^2 \tanh^2(u y)} \label{Idefn} 
\eea
As $H$ and $I$ are symmetric about the origin, it is sufficient to determine if $\int_0^\infty dx H(x) e^{I(x)}$ is finite. For $u > 0$ and large, positive $x$, we can write
\bsn
{H(x) \stackrel{x \gg 1}{\sim}\label{Hasymp}} 
\frac{2 v_0}{v_0^2-v_1^2} ~&,~ $v_0 \neq v_1$ \\
\frac{e^{2 u x}}{2} ~&,~ $v_0=v_1$
\esn
and 
\bsn
{I(x) \stackrel{x \gg 1}{\sim} \label{Iasymp}} 
-\frac{2(\gamma_0 v_1+\gamma_1 v_0 (1-\delta_{g,0}))}{v_0^2-v_1^2} x ~&,~ $v_0 \neq v_1$ \\
-\frac{\gamma_0+\gamma_1 (1-\delta_{g,0})}{4 u v_0} e^{2 u x} ~&,~ $v_0=v_1$
\esn
which show that $e^{I(x)}$ decays much faster than $H(x)$ grows at large $x$ and therefore, the normalization constant (\ref{norm}) is finite. For $u=0$, $H=\frac{2}{v_0}, I=-\frac{2 \gamma_1}{v_0} \int_0^x dy \tanh(g y) \stackrel{ x \gg 1}{\to}-\frac{2 \gamma_1}{v_0}x (1-\delta_{g,0})$ and therefore, the stationary state exists if $g > 0$. The above discussion also shows that the steady distribution is always unbounded and has exponential or double-exponential tail.

\begin{figure}[t]
\centering
\includegraphics[scale=0.52]{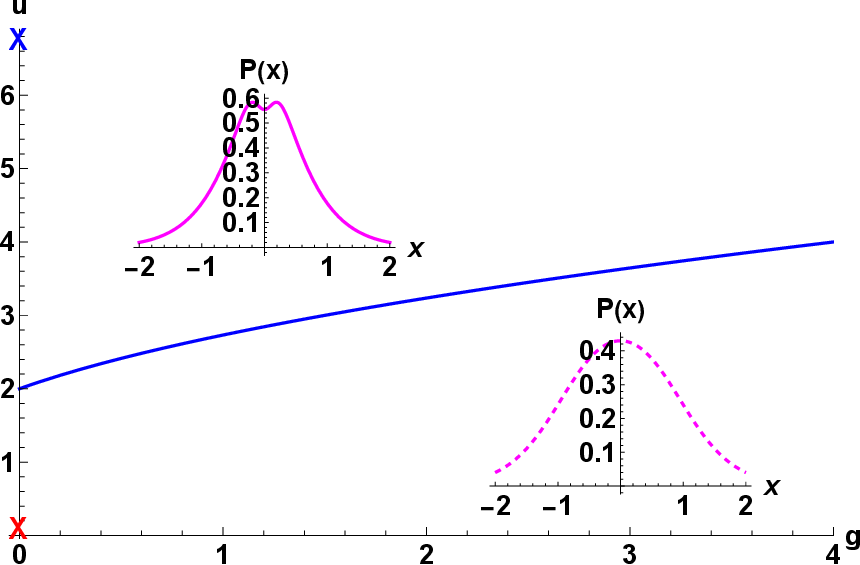}
\includegraphics[scale=0.48]{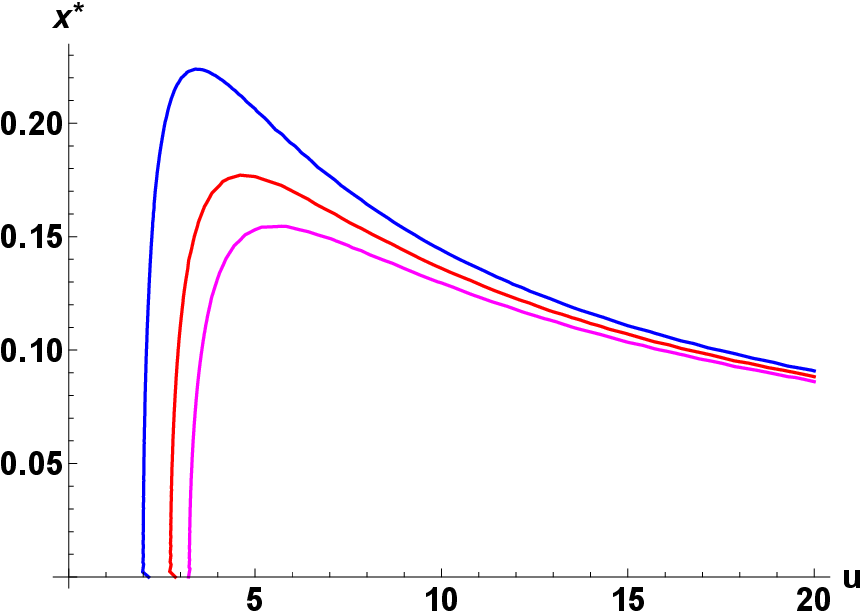}
\caption{(a) Steady state distribution for $\gamma_0=v_0=1, \gamma_1=v_1=1/2, g=1/2$, and $u=1$ (dashed)  and $u=5$ (solid) where the (blue) curve in $u-g$ plane shows (\ref{quadcond}). The $\times$ mark in red at the origin signifies that the stationary distribution does not exist for $u=g=0$, and the $\times$ mark in blue stands for $g=0, u \to \infty$ to show that the distribution is unimodal in this limit (see Sec.~\ref{equil} for details). (b) The peak of the bimodal distribution occurs at the position, $\pm x^*$ and is given by (\ref{modecond}). Figure shows the magnitude $|x^*|$ as a function of $u$ for $g=0$ (blue), $1$ (red) and $2$ (magenta). The rest of the parameters are $\gamma_0=v_0=1, \gamma_1=v_1=1/2$.}
\label{modefig}
\end{figure}

We now consider the small-$x$ behavior of $P(x)$; on expanding the function $H(x)$ in (\ref{Hdefn}) to quadratic order and the integral $I(x)$ in (\ref{Idefn}) to linear order in $x$ about the origin, we find that 
\be
\frac{P(x)}{P(0)} \stackrel{x \to 0}{\approx} 1-\frac{g \gamma_1 v_0+u v_1 (\gamma_0-u v_1)}{v_0^2} x^2
\ee
 which means that the distribution has a minimum at the origin  if 
 \be
 v_1^2 u^2 -\gamma_0 v_1 u -g \gamma_1 v_0 > 0 \label{modebdry}
 \ee
and suggests that the distribution can be multimodal. On taking the derivative of (\ref{ssexact}) with respect to $x$ and setting it equal to zero, we find that the stationary state distribution has extrema at $x^*$ that obey the following exact condition: 
\bea
\cosh^2(u x^*) [\gamma_1 v_0 \tanh (g x^*)+ \gamma_0 v_1 \tanh (u x^*)]  &=& u v_1^2 \tanh (u x^*) 
\label{modecond}
\eea
The above equation shows that $x^*=0$ is always a mode of the stationary state distribution and for $u=0$, the stationary distribution has a single mode; unfortunately, it does not seem possible to find a closed form expression for $x^*$ for arbitrary $u, g$ except for some special parameters in the $u-g$ plane. These special cases (discussed in Sec.~\ref{spl}) suggest that (\ref{modecond}) has at most three solutions and the parameters where the transition between the unimodal and bimodal distribution occurs is consistent with the criterion (\ref{modebdry}). 
Therefore, if we now assume that there are at most two maxima, from (\ref{modebdry}), we find that the distribution is bimodal if
 \be
 (u - u^*_+) (u-u^*_-) > 0 ~,~u_\pm^*=\frac{\gamma_0 \pm \sqrt{\gamma_0^2+4 g \gamma_1 v_0}}{2 v_1}
 \ee
 and unimodal otherwise.  For $u, g > 0$, as $u_-^* < 0$, the distribution has two maxima if 
 \be
 u > u_+^*=\frac{\gamma_0 + \sqrt{\gamma_0^2+4 g \gamma_1 v_0}}{2 v_1} , v_1 \neq 0 \label{quadcond}
 \ee
 as illustrated in Fig.~\ref{modefig}a. 

\subsection{Special cases}
\label{spl}

As described in Appendix~\ref{app_ss}, except for $u \to \infty, v_0=v_1$, the steady state distribution $P_\pm(x)$ of the right- and left-movers is given by
 \bea
 P_+(x) &=& P_+(0) e^{\int_0^x \left( \frac{\gamma_-(x')}{v_-(x')}- \frac{\gamma_+(x')}{v_+(x')} -\frac{v'_+(x')}{v_+(x')}\right) dx'} \label{sspexact} \\
 P_-(x) &=&P_-(0) e^{\int_0^x \left(\frac{\gamma_-(x')}{v_-(x')}- \frac{\gamma_+(x')}{v_+(x')}-\frac{v'_-(x')}{v_-(x')} \right) dx'} \label{ssnexact} 
\eea
 where $P_\pm(0)$ are obtained using  $\int_{-\infty}^\infty dx' P_\pm(x')=1/2$. 
 Using (\ref{vchoice}) and (\ref{gchoice}) in the above equations and in (\ref{ssexact}) for the total distribution, $P(x)=P_+(x)+P_-(x)$, a general expression for the stationary state can be obtained but it involves Gauss hypergeometric function and is not particularly illuminating. However, simple expressions for the stationary distribution can be obtained for following special points in the parameter space: $g > 0, u=0$ [Sec.~\ref{tumble}], $g=0, u > 0$ [Sec.~\ref{equil}] and $u=g > 0$ [Sec.~\ref{tumblee}].


\subsubsection{Uniform speed, variable tumbling rates}
\label{tumble}

For constant speed but inhomogeneous tumbling rates, $v_\pm=v_0, \gamma_\pm(x)=\gamma_0 \pm \gamma_1 \tanh(g x)$ with $g >0$, using (\ref{ssexact}), (\ref{sspexact}), (\ref{ssnexact}), we find the following simple expressions, 
\bea
P_+(x) &=& \frac{g \Gamma \left(\frac{1}{2}+ \frac{ \gamma_1}{g
   v_0}\right)}{2\sqrt{\pi } \Gamma \left(\frac{\gamma_1}{g
   v_0}\right)} \left[\cosh(g x) \right]^{-\frac{2 \gamma_1}{g v_0}} \label{v1zerop} \\
P_-(x) &=& P_+(x)\\
P(x) &=& 2 P_+(x) \label{v1zero}
\eea
which are independent of $\gamma_0$. We verify that the above distributions are unimodal in agreement with (\ref{modebdry}) for $u=0$ and the tail of the distribution is exponential, $P(x) \stackrel{g |x| \gg 1}{\sim} \frac{\gamma_1}{v_0} e^{-\frac{2 \gamma_1 |x|}{v_0}}$ (see also (\ref{v1zerostep}) below). 

In later discussion in Sec.~\ref{sec:dyn}, we will be interested in the limit $g \to \infty$ where 
\be
\gamma_\pm(x)=\gamma_0 \pm \gamma_1 \textrm{sgn}(x) \label{ginflim}
\ee
so that the tumbling rate is discontinuous at $x=0$. In this case, (\ref{v1zero})  reduces to 
\be
P(x)=\frac{\gamma_1}{v_0} e^{-\frac{2 \gamma_1 |x|}{v_0}} ~~ (u=0, g \to \infty) \label{v1zerostep}
\ee
which is continuous at the origin but its first derivative is discontinuous at $x=0$, and gives the variance in the particle's position to be 
\be
\sigma^2=\int_{-\infty}^\infty dx x^2 P(x)=\frac{v_0^2}{2 \gamma_1^2} ~~ (u=0, g \to \infty) \label{varu0ginf}
\ee
We also find that the variance for finite $g$ decreases with increasing $g$, and approaches (\ref{varu0ginf}) as $g^{-1}, 0< g \ll 1$ and $g^{-2}, g \gg 1$. In fact, we will see in Sec.~\ref{sec:dyn} that the full dynamics of the variance for finite $g$ are well approximated by those for $g \to \infty$ for $g >1$.


\subsubsection{Variable speed, uniform  tumbling rates}
\label{equil}

\begin{figure}[t]
\centering
\includegraphics[scale=0.5]{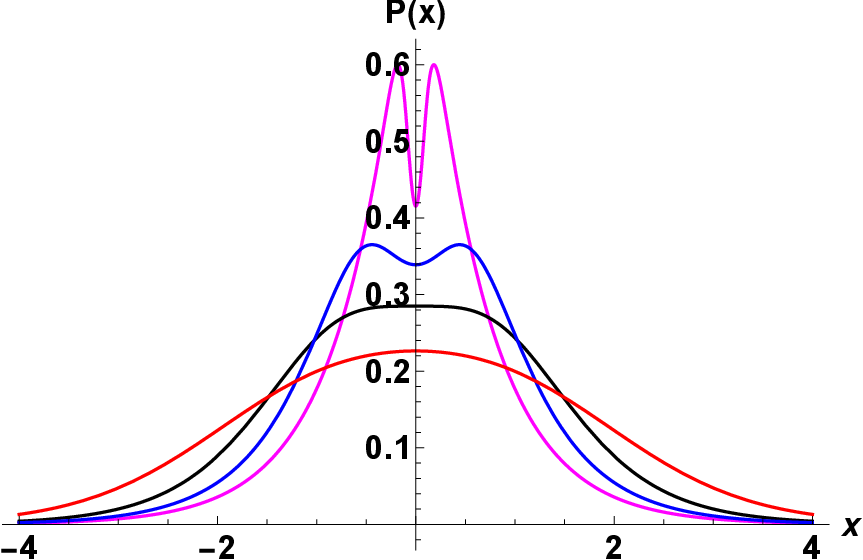}
\includegraphics[scale=0.5]{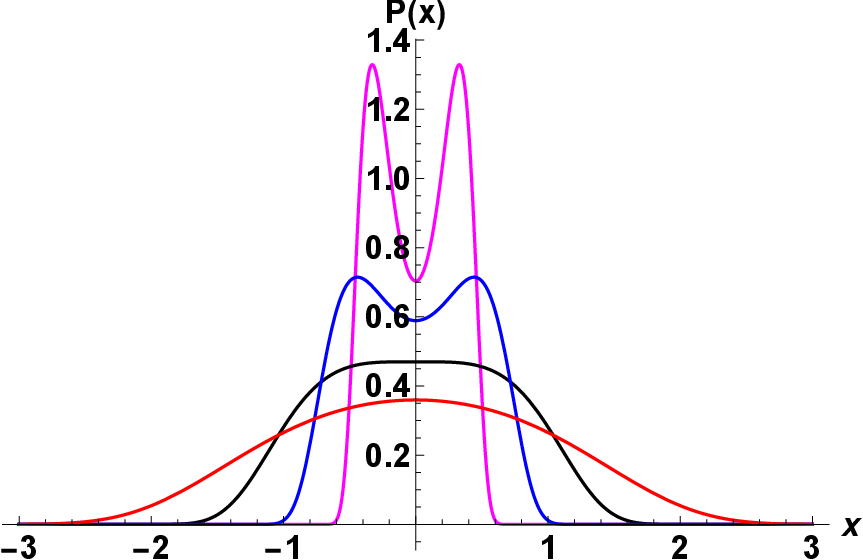}
\caption{Variable speed, uniform  tumbling rates: Stationary state distribution for (a) $v_0=3/2, v_1=1, \gamma_0=1, g=0$ and $u=1/2$ (red), $1$ (black), $2$  (blue), and $10$ (magenta) and (b) $v_0=v_1=1, \gamma_0=1, g=0$ and $u=1/2$ (red), $1$ (black), $2$  (blue), and $4$ (magenta). Due to (\ref{modeg0}), the transition between unimodal and bimodal distribution occurs at $u=1$ for both set of parameters.}
\label{distg0}
\end{figure}  

For constant  tumbling rates but inhomogeneous speed, $v_\pm=v_0 \mp v_1 \tanh(u x), \gamma_\pm(x)=\gamma_0$, we obtain 
\bea
P_\pm(x) &\propto&  \frac{1}{v_0 \mp v_1 \tanh(u x)} \exp \left[-\frac{\gamma_0 v_1}{u \left(v_0^2-v_1^2\right)} \ln \left(\cosh ^2(u x)-\frac{v_1^2 \sinh ^2(u x)}{v_0^2}\right) \right] \label{eq:ppm}    \\ 
P(x) & \propto & \frac{1}{v_0^2-v_1^2 \tanh^2(u x)} \exp \left[-\frac{\gamma_0 v_1}{u \left(v_0^2-v_1^2\right)} \ln \left(\cosh ^2(u x)-\frac{v_1^2 \sinh ^2(u x)}{v_0^2}\right) \right]
\label{equil2}
\eea
when $v_0 \neq v_1$. At large distances from the origin, these distributions decay exponentially fast and the decay rate is independent of $u$, $P(x) \stackrel{u |x| \gg 1}{\sim} e^{-\frac{2 \gamma_0 v_1 |x|}{ v_0^2-v_1^2}}$. 
For $v_0=v_1$, we have 
\bea
P_\pm(x) &\propto& \frac{1}{1 \mp \tanh(u x)} e^{-\frac{\gamma_0 \sinh ^2(u x)}{u v_0}} \\
P(x) &\propto& \cosh ^2(u x) e^{-\frac{\gamma_0 \sinh ^2(u x)}{u v_0}} 
\label{v01g0}
\eea
which show that the distributions decay super-exponentially, $P(x) \stackrel{u |x| \gg 1}{\sim} e^{-\frac{\gamma_0 e^{2 u |x|}}{4 u}}$. This rapid decay is a consequence of the fact that the right-mover's (left-mover's) speed approaches zero as $x \to \infty$ ($x \to -\infty$) and therefore, the particle can not be found at large distances.

For $g=0$, the condition (\ref{modecond}) for multimodality  simplifies, and the location of the modes is determined by $\tanh(u x^*)=0$ or $\cosh ^2(u x^*)=\frac{u v_1}{\gamma_0}$ which immediately gives 
\bsn
{x^*= \label{modeg0}}
0 ~&,~ $0 < u < \frac{\gamma_0}{v_1}$ \\
0, u^{-1}\cosh^{-1} \left(\pm \sqrt{\frac{u v_1}{\gamma_0}} \right)~&,~ $u > \frac{\gamma_0}{v_1}$
\esn
so that the distribution $P(x)$ is unimodal for $0 < u < \frac{\gamma_0}{v_1}$ and is bimodal otherwise with the minimum at the origin. The transition curve $u=\frac{\gamma_0}{v_1}$ between unimodal and bimodal distribution is  independent of $v_0$ and is also obtained from (\ref{quadcond}). Note that for $u > \frac{\gamma_0}{v_1}$, the location $|x^*|$ of the maxima is a non-monotonic function of $u$ as shown in Fig.~\ref{modefig}b, which initially increases as $\sqrt{u-\frac{\gamma_0}{v_1}}$ and then decays to zero, $|x^*| \stackrel{u \gg 1}{\sim} \frac{\ln u}{u}$ [also see (\ref{uinf}) below]. The distribution $P(x)$ of the position of the particle is shown in Fig.~\ref{distg0} for various values of $u$, and describes the features discussed above.

\begin{figure}[t]
\centering
\includegraphics[scale=0.5]{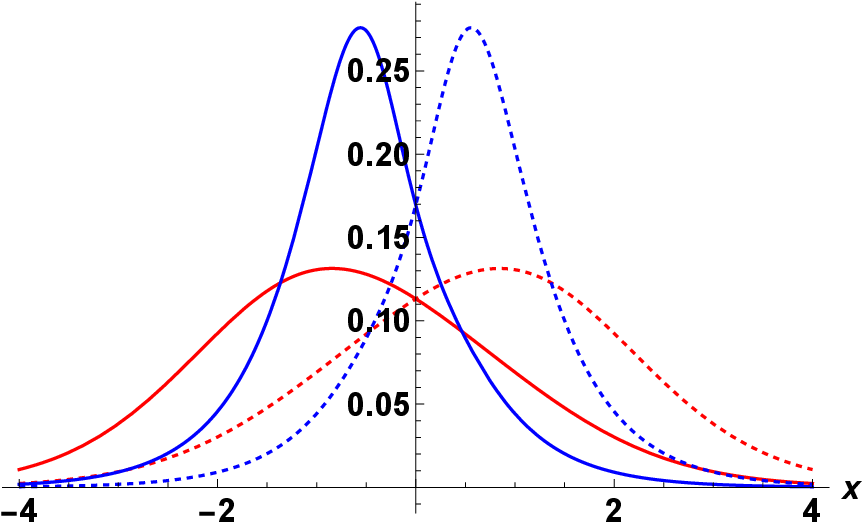}
\includegraphics[scale=0.5]{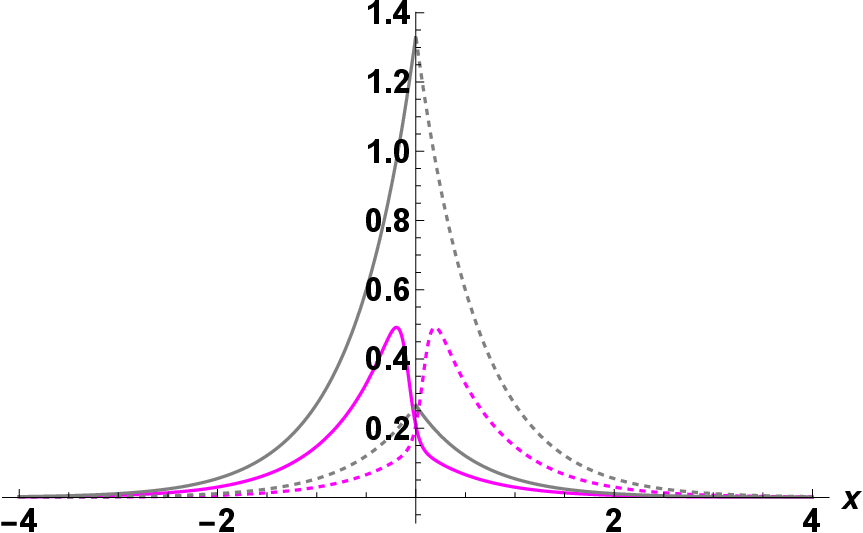}
\caption{Variable speed, uniform  tumbling: Stationary state distribution $P_+(x)$ (dashed) and $P_-(x)$ (solid) for $v_0=3/2, v_1=1, \gamma_0=1, g=0$ when (a) $u=1/2$ (red) and $2$  (blue), and (b) $u=10$ (magenta) and $u \to \infty$ (gray).}
\label{distv1zrl}
\end{figure}  

Figure \ref{modefig}a shows that for a given $g \geq 0$, as one increases $u$, the steady state position distribution changes its shape from unimodal to a bimodal one. To physically understand why bimodality appears for large $u$, we recall that for $g=0$, the active particle experiences a confining potential $V(x)$ given by (\ref{potential}) which has a minimum at the origin. Both left-movers and right-movers cross the origin with speed $v_0$ which allows the particle to traverse some distance from the origin until the potential cost makes it difficult for the particle to go further. Indeed, as Fig.~\ref{distv1zrl} shows, the distributions $P_+(x)$ and $P_-(x)$ display a peak at positive and negative $x$, respectively. If now $u$ increases, the potential becomes steeper which makes it difficult for the particle to traverse larger distances after crossing the origin; as a result, the distributions $P_\pm (x)$ become narrower and their peak occurs closer to the origin - see Fig.~\ref{distv1zrl}. Thus $P(x)=P_+(x)+P_-(x)$ shows a dip at the origin because both $P_+(x)$ and $P_-(x)$ fall sharply from their peak value near the origin when $u$ is large. Interestingly, the decrease in width of $P_\pm (x)$ and the shifting of peak position towards the origin with increasing $u$ also suggests that for $u \to \infty$, the distributions $P_\pm (x)$ reach their maximum value at the origin and also develop a finite discontinuity at the origin. Below we discuss this case more explicitly.

For $u \to \infty$, the speed $v_\pm(x)=v_0 \mp v_1 \text{sgn}(x)$ changes discontinuously at the origin, and the stationary state distribution is given by
\bsn
{P(x) = \label{uinf}}
\frac{\gamma_0 v_1}{ v_0^2-v_1^2} e^{-\frac{2 \gamma_0 v_1 |x|}{ v_0^2-v_1^2}} ~&,~ $v_0 \neq v_1$\\
\delta(x)  ~&,~$v_0 = v_1$ \label{splcase}
\esn
(see Appendix~\ref{app_ss} for (\ref{splcase})). The position distribution thus collapses from a non-localized distribution for $v_0 \neq v_1$ to a localized one at the origin for $v_0=v_1$; this result has previously been obtained in \cite{Dhar:2019} that describes a run-and-tumble particle with uniform tumbling rate and moving in a confining potential $V(x) \sim |x|$, and corresponds to $u \to \infty, g=0$ in our model. Note that the large-$x$ behavior of $P(x)$ for finite $u$ is identical to that in $u \to \infty$ limit; this is simply because for $|x| \gg u^{-1}$, the inhomogeneous speed becomes constant and behaves the same way as for $u \to \infty$. 
For $v_0 \neq v_1$, the distribution $P(x)$ is continuous but not differentiable at $x=0$, that is, there is a finite discontinuity in the first derivative at the origin. But the distribution of right- and left-movers is given by 
\bea
P_\pm(x) &=& \frac{\gamma_0 v_1}{ 2 v_0} \frac{e^{-\frac{2 v_1 \gamma_0 |x|}{v_0^2-v_1^2}}}{v_0\mp v_1 \textrm{sgn}(x)}
\eea
which shows that these distributions have a finite discontinuity at the origin,  $|P_\pm(x\to 0+)-P_\pm(x \to 0-)|=\frac{2 \gamma_0 v_1^2}{v_0 (v_0^2-v_1^2)}$ (see Fig.~\ref{distv1zrl}b also). This behavior is unlike that for $u=0, g \to \infty$ limit (see Sec.~\ref{tumble}) where the distributions $P_\pm(x)$ are seen to be continuous at the origin.

Finally, we remark that as mentioned in Sec.~\ref{ch2}, for $g=0$, the RTP is an active particle in an external potential $V(x)$ given by (\ref{potential}). But for $v_0, \gamma_0 \to \infty$ with the diffusion constant, $2 D=v_0^2/\gamma_0$ finite, one expects the above distributions reduce to their equilibrium or passive counterpart \cite{Sevilla:2019,Dhar:2019}. On taking these limits in (\ref{equil2}), we verify that indeed
\be
P(x) \propto e^{-\frac{V(x)}{D}}= [\cosh(u x)]^{-\frac{1}{u D}}
\ee


\subsubsection{Inhomogeneous speed and tumbling rates}
\label{tumblee}

 For $g=u > 0$ in (\ref{vchoice}) and (\ref{gchoice}), the stationary state distribution (\ref{ssexact}) is given by 
 \bea
P(x)  &\propto&   \frac{1}{v_0^2-v_1^2\tanh ^2(u x)} \exp \left[-\frac{\gamma_0 v_1+\gamma_1v_0}{u \left(v_0^2-v_1^2\right)} \ln \left(\cosh^2(u x) -\frac{v_1^2 \sinh^2(u x)}{v_0^2}\right)\right]
\eea  
which reduces to  (\ref{equil2}) for $g=0$.  
The  above distribution also has a transition between unimodal and bimodal shapes which is found using either (\ref{modecond}) or (\ref{quadcond}) at $u=\frac{\gamma_0 v_1+\gamma_1 v_0}{v_1^2}$. 
As in Sec.~\ref{equil}, the distribution decays super-exponentially for $v_0=v_1$ and exponentially fast otherwise, and the location of the maxima is a non-monotonic function of $u$ (see Fig.~\ref{modefig}b). For $u=g \to \infty$, the speeds and the tumbling rates are discontinuous at the origin, and we obtain the stationary distribution in this case on replacing $\gamma_0 v_1 $ in  (\ref{uinf}) by $\gamma_0 v_1+\gamma_1v_0$ .

\section{Dynamics far from steady state}
\label{sec:dyn}

In this section, we focus on the dynamics of the system when the particle is initially at position $x_0$ (see (\ref{initialcond})) and, without loss of generality, assume that $x_0 \geq 0$. We are interested in the behavior of the mean-squared displacement (MSD) denoted by $\sigma^2(x_0, t)$ as a function of time. While in the limit  $t \to \infty$, the MSD reaches the stationary state result, $\sigma^2=\int_{-\infty}^\infty dx x^2 P(x)$ where $P(x)$ is given by (\ref{ssexact}) and is independent of $x_0$, our Monte Carlo simulations show that at short and intermediate times, $\sigma^2(x_0, t)$ displays interesting features depending on the initial position of the particle. We find at short times, the MSD grows with time as a power law for any initial position but at intermediate times,  if $x_0$ is sufficiently large, it shows a plateau or a peak before finally relaxing exponentially fast to the stationary state.

To get an analytical insight into these behavior, we note that as long as the particle remains sufficiently far on the right side of the origin, the  dynamics are governed by the homogeneous limit in which the rates (\ref{vchoice}) and (\ref{gchoice}) saturate to a constant: 
\bea
v_\pm(x \to \infty) &=& v_0 \mp v_1 \label{univ}\\
\gamma_\pm(x \to \infty)  &=& \gamma_0 \pm \gamma_1. \label{unig}
\eea
In Appendix \ref{app_cnst}, we obtain exact expressions for the mean position and MSD for the above homogeneous rates (note that for these homogeneous rates, the case of uniform speed or tumbling is obtained by setting $v_1=0$ or $\gamma_1=0$, respectively). In the following, we explain in detail how these results describe the dynamics at short and intermediate times for the general model in Sec.~\ref{superdiff} and Sec.~\ref{peaksec}, respectively. However, once the particle is sufficiently close to the origin or if its initial position is small enough, it becomes sensitive to the inhomogeneity in rates and the above homogenous rates approximation is no longer valid, and we describe the long-time dynamics in Sec.~\ref{exposec} using simulations. Although we have not been able to capture the full dynamics analytically for general parameters, we obtain some exact results when $u=0, g \to \infty$ which are described in Appendices~\ref{app_Green}-\ref{app_msdn} and are also discussed below.  

\subsection{Super-diffusive growth at short times}
\label{superdiff}

\begin{figure}[t]
\centering
\includegraphics[scale=0.6]{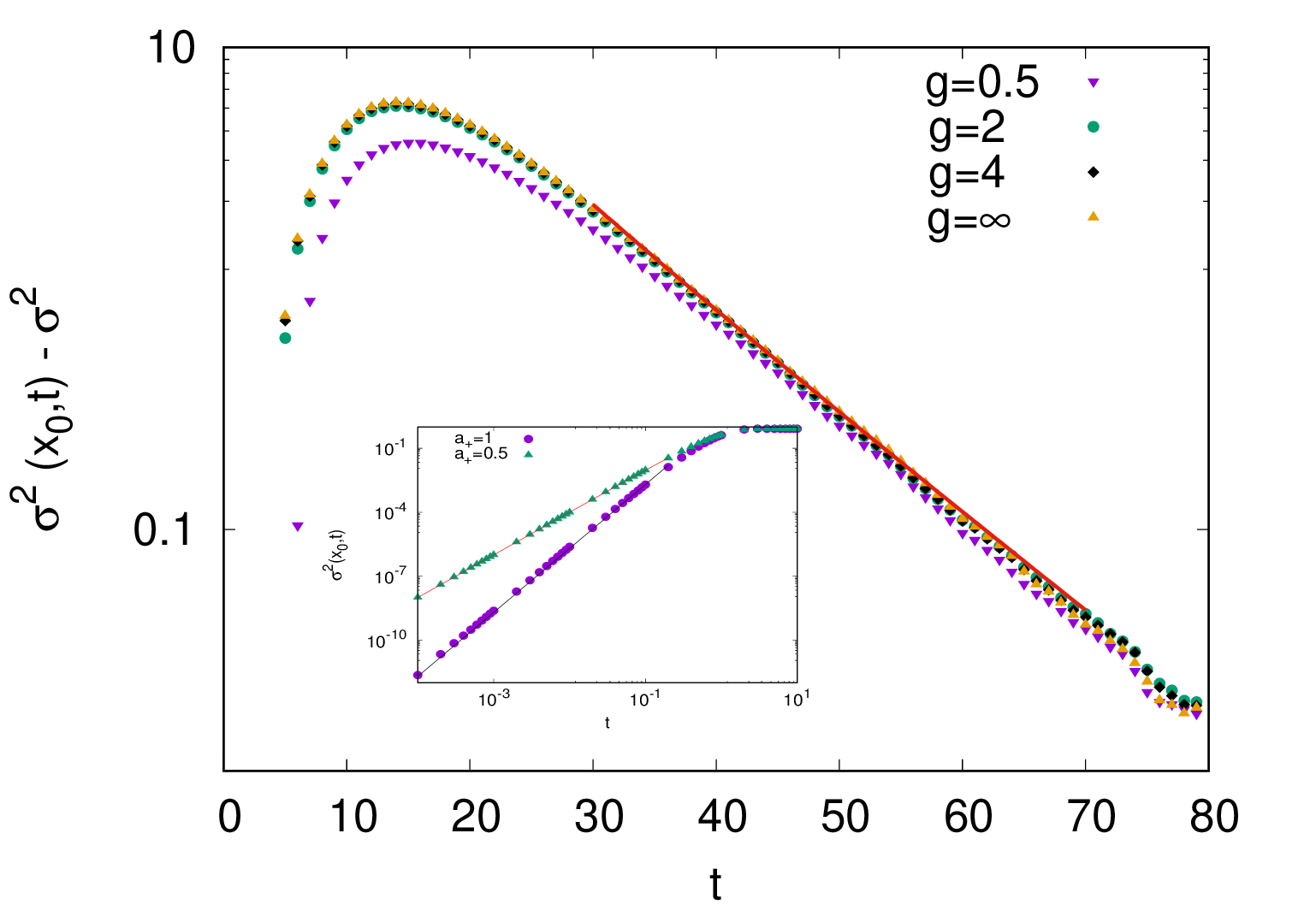}
\caption{Uniform speeds, variable tumbling rates: The main figure shows the deviation of the mean-squared displacement (MSD) $\sigma^2(x_0,t)$ from the stationary state result $\sigma^2$ for $\gamma_0 \neq \gamma_1$. The simulation data (points) and the analytical prediction (\ref{expo1}) shown by solid line are obtained for $u=0$, $x_0=8$, $\Delta =0$, $v_0=1$, $\gamma_0=1$, $\gamma_1=0.4$ and various different $g$ values shown in figure legends. Equation (\ref{eq:tstar}) gives the peak time $t^\ast = 16.25$ which is close to the observed peak time. The inset shows the superdiffusive growth of MSD at short times obtained using simulations (points) and (\ref{varFs}) (solid line) for $u=0$, $g=1$, $x_0=1$, $v_0=1$ and $a_+=1,0.5$.}
\label{fig:u0}
\end{figure}

As shown in the inset of Fig.~\ref{fig:u0}, at short enough times, the MSD grows as a power law; 
this is a generic result and holds for any run-and-tumble motion, including the case when velocities and tumbling rates are uniform, and can be seen as follows.  
For the homogeneous rates (\ref{univ}) and (\ref{unig}), from  (\ref{cstmean}), we find the mean displacement to be
\be 
{\bar x}_{homo}(t) \approx x_0+(\Delta v_0-v_1) t, \;\;\;\;\;\; \mbox{$t \ll (2\gamma_0)^{-1}$} \label{meanFs}
\ee
where $\Delta=a_+ - a_-$. Also, from (\ref{cstvar}), we find that the MSD is given by  
\be 
\sigma_{homo}^2(t) \approx (1-\Delta^2) v_0^2 t^2+\frac{2}{3} [(3 \Delta^2-1) \gamma_0+2 \Delta \gamma_1] v_0^2 t^3, \;\;\;\;\;\; \mbox{$t \ll (2\gamma_0)^{-1}$} \label{varFs}
\ee
which reduces to that obtained in  \cite{Jose:2023} for symmetric, homogeneous rates and fully asymmetric initial condition on setting $v_1=\gamma_1=0, \Delta=1$. The above expression (\ref{varFs}) shows that the MSD grows with time as a power law with an exponent $2$ for $0<a_+ <1$, while for $a_+=1$ and $a_+=0$, the growth law is $t^3$. Our simulation data presented in the inset of Fig. \ref{fig:u0} is seen to be in excellent agreement with (\ref{varFs}) for times much smaller than the typical time scale for tumbling. On setting $v_1=0$ in (\ref{meanFs}) and (\ref{varFs}), we find that the resulting expressions also agree with the exact results obtained in Appendix~\ref{app_msd0} for $u=0, g \to \infty, \gamma_0=\gamma_1$ when $2 \gamma_0 t \ll 1$.

\begin{figure}[t]
\centering
\includegraphics[scale=0.6]{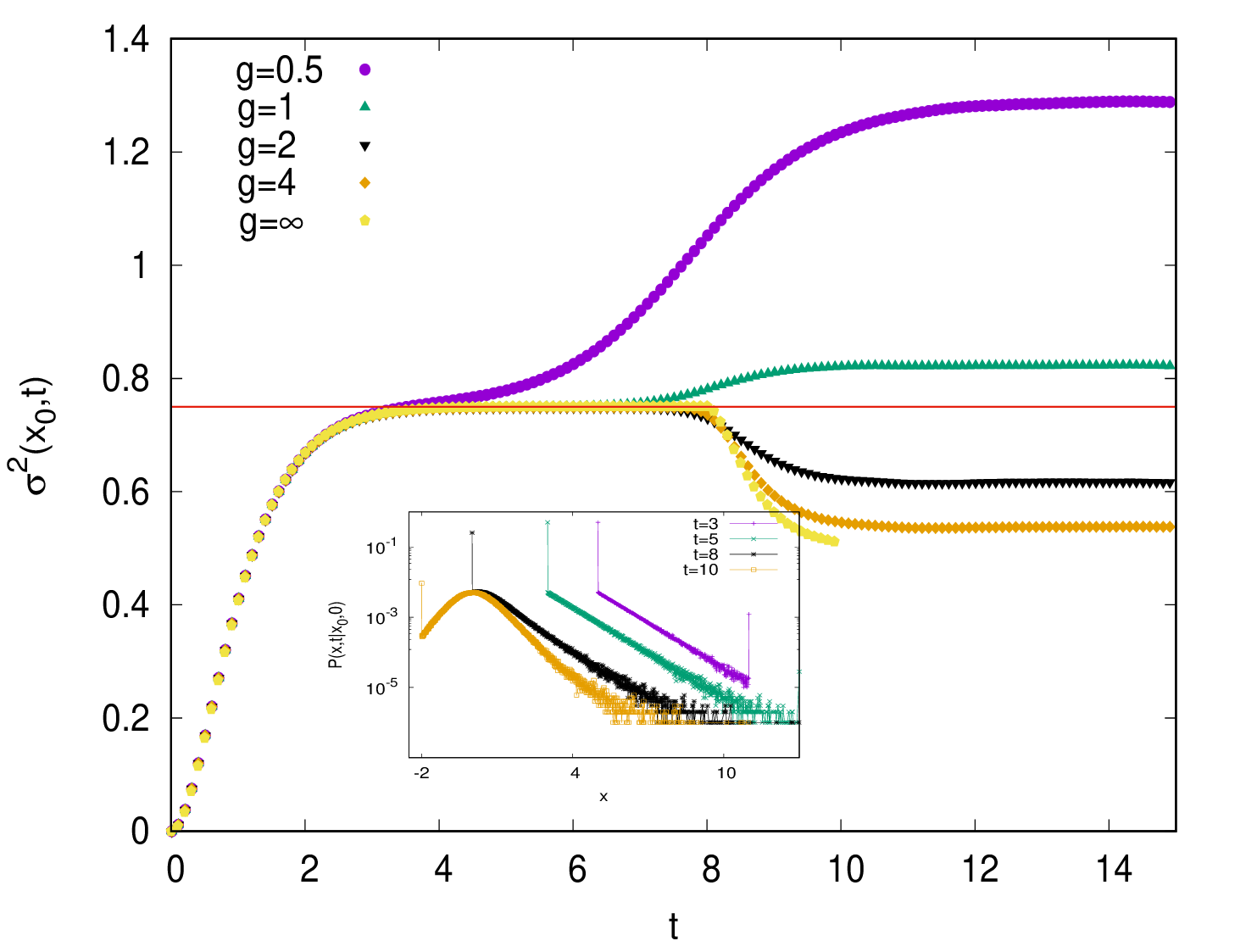}
\caption{Uniform speeds, variable tumbling rates: The main figure shows the mean-squared displacement (MSD) $\sigma^2(x_0,t)$ for $\gamma_0=\gamma_1$. The  data shown are for the same parameters as in Fig.~\ref{fig:u0}, except $\gamma_0=\gamma_1=1$ here. The solid line shows the analytical prediction (\ref{constsig}) for the plateau value, and the theoretical value of $t^\ast = x_0/v_0=8$ matches with large $g$ data. The inset shows the distribution $P(x,t|x_0,0)$ for $g=1$ at different time slices as mentioned in the legend.}
\label{fig:u1}
\end{figure}

\subsection{Plateau or peak at intermediate times}
\label{peaksec}

As shown in the inset of Fig.~\ref{fig:u0}, for small $x_0$, the MSD increases monotonically towards the stationary 
state - at early times, it increases algebraically,  as already explained in Sec.~\ref{superdiff} and then it approaches the steady state value exponentially fast, as described later in Sec.~\ref{exposec}. However, for large $x_0$, we observe an intermediate time regime, where after the initial super-diffusive growth and before the late time relaxation, the MSD displays a plateau for $\gamma_0 = \gamma_1$ or a peak for  $\gamma_0 > \gamma_1$.

\begin{figure}[t]
\centering
\includegraphics[scale=0.6]{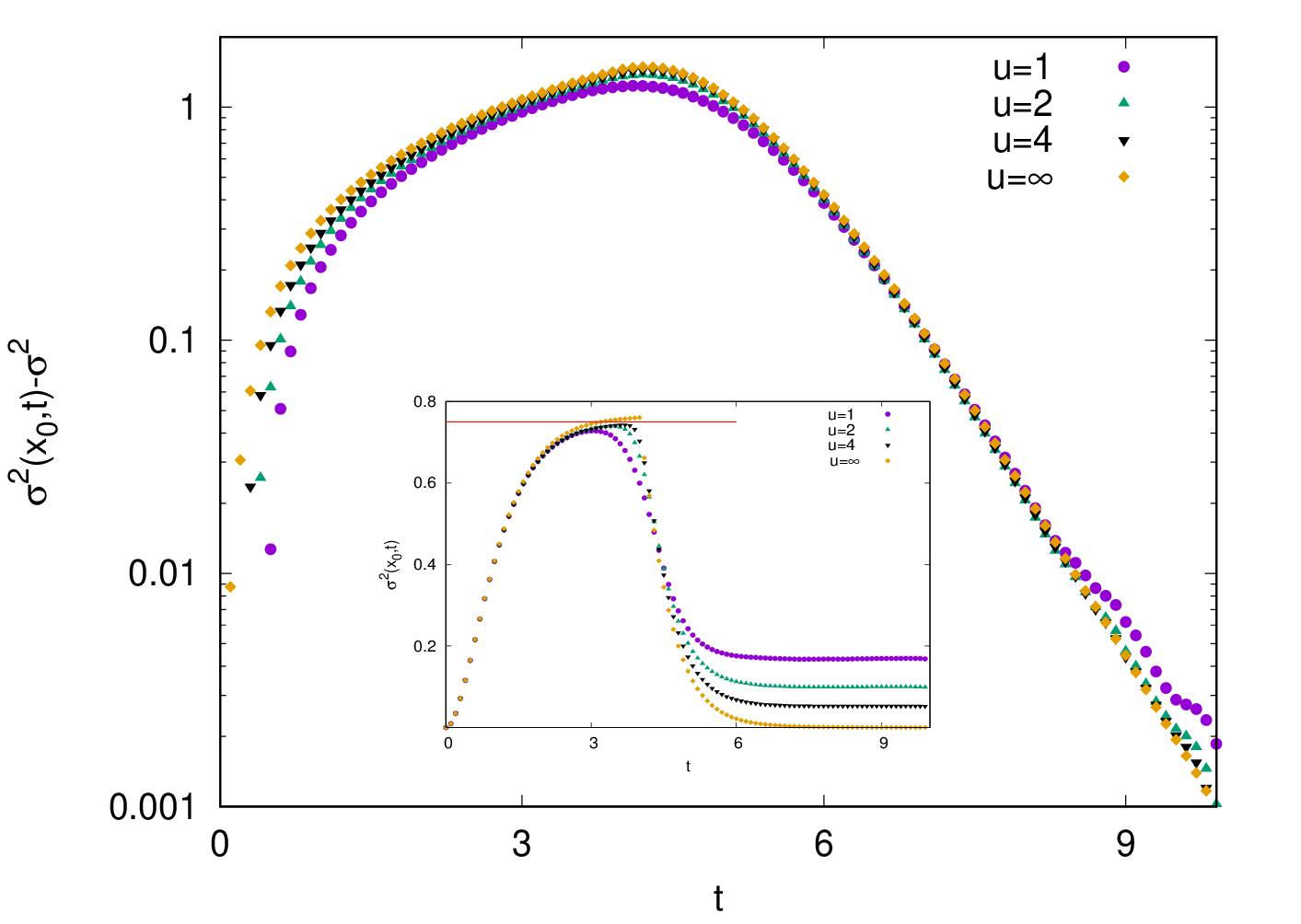}
\caption{Variable speeds, variable tumbling rates: The main figure shows the mean-squared displacement (MSD) $\sigma^2(x_0,t)$ for different $u$ values as shown in the legend; the other parameters are $v_0=1$, $v_1=1$, $\gamma_0=2$, $\gamma_1=1$, $g=4$, $x_0=8$ and $\Delta =0$. The inset shows the plateau in MSD for $\gamma_0=\gamma_1=1$, while all the other parameters remain the same as in the main plot. The solid line in the inset corresponds to the predicted level of plateau from (\ref{varFl}) which shows reasonable agreement for large $u$.}
\label{fig:ug4}
\end{figure}

For $\gamma_0 = \gamma_1$, as shown in Fig.~\ref{fig:u1} for $u=0$ and various values of $g$ and the inset of Fig. \ref{fig:ug4} for nonzero $g$ and different values of $u$, the MSD reaches a plateau over an intermediate time range. 
The tumbling rate $\gamma_-(x)$ of left-mover vanishes at large distances from the origin which means that far from the origin, once the particle turns leftward, it does not tumble anymore and only moves ballistically with a finite speed towards left. For such deterministic motion, the MSD which is a measure of fluctuations, does not increase with time in the intermediate time regime. To see  this effect more explicitly, in the inset of Fig. \ref{fig:u1}, we show the conditional probability distribution $P(x,t|x_0,0)$ obtained from simulations as a function of $x$ for some time slices. We find that $P(x,t|x_0,0)$ varies exponentially with $x$ and shows two delta-function peaks at the two edges, which correspond to the persistent trajectories where the particle did not tumble even once. At short times, the peak at the left edge of the distribution does not diminish with time, which is consistent with zero tumbling rate. But at large times, the peak height starts decreasing again when the tumbling rate becomes nonzero.

The behavior described above can be quantitatively captured by the homogeneous rates (\ref{univ}) and (\ref{unig}); from (\ref{cstmean}) and (\ref{cstvar}), it follows that for $t \gg (2 \gamma_0)^{-1}$, the mean position and MSD of a particle  moving with these rates are given by 
\be 
{\bar x}_{homo}(t) \approx x_0+\frac{v_0 (\Delta \gamma_0+\gamma_1)}{2 \gamma_0^2}-\frac{t (\gamma_0 v_1+\gamma_1 v_0)}{\gamma_0}, \;\;\;\;\;\; \mbox{$t \gg (2\gamma_0)^{-1}$} \label{meanFl}
\ee
and 
\be 
\sigma_{homo}^2(t) \approx \frac{(\gamma_0^2-\gamma_1^2) v_0^2 t}{\gamma_0^3}+\frac{(5 \gamma_1^2-2 \gamma_0^2-\Delta^2 \gamma_0^2+2 \Delta \gamma_0 \gamma_1) v_0^2}{4 \gamma_0^4}, \;\;\;\;\;\; \mbox{$t \gg (2\gamma_0)^{-1}$}. \label{varFl}
\ee
Then it follows from (\ref{varFl}) that for $\gamma_0 = \gamma_1$, the MSD indeed becomes time-independent and stays constant at a value 
\be
\sigma_{homo}^2(t)  \stackrel{\gamma_0=\gamma_1}{=}\dfrac{(3-\Delta^2 + 2 \Delta)v_0^2}{4 \gamma_0^2} \label{constsig}
\ee
Note that although the homogeneous rates do not allow the system to reach a steady state, the variance in the particle's position becomes time-independent for the reason described above.  We compare (\ref{constsig}) with our simulations in the main plot of Fig. \ref{fig:u0} and the inset of \ref{fig:ug4}, and find a good agreement. 

As a stationary state exists for the rates (\ref{vchoice}) and (\ref{gchoice}), we expect that there is a time scale $t^*$ that marks the end of the plateau in the MSD and the beginning of the relaxation towards the steady state value. To obtain an insight into this time scale, we consider the limit of $u=0, g \to \infty$ which means that the particle moves with a uniform velocity and the tumbling rates vary in a step-function manner. For this case, using the exact Green's function given by (\ref{finalG}) in Appendix \ref{app_Green}, we calculate the mean and variance of the position of the particle  as a function of time in Appendix \ref{app_msd0}. We find that the relaxation regime sets in over a time scale $t^*=\max \left ( \dfrac{x_0}{v_0}, \dfrac{1}{2 \gamma_0} \right )$ which may be understood as follows: we first note that for $2 \gamma_0 t \gg 1$, the mean position (\ref{meaneqinter}) of the particle coincides with (\ref{meanFl}) when $v_1=0$ and $\gamma_0=\gamma_1$ which means that the dynamics can be described by the homogeneous rates model during the intermediate time regime. But the latter model is not valid when the mean position ${\bar x}_{homo}(t)$ vanishes as the inhomogeneous tumbling rates (\ref{ginflim}) change their behavior at the origin. This time is given by  $ \dfrac{x_0}{v_0} + \dfrac{(1+\Delta)}{2 \gamma_0}$ and  is consistent with the above expression for $t^*$.

To reiterate, for $u=0, g \to \infty$, the time scale $t^\ast$ denotes the time at which the mean position of the particle moving with homogeneous rates becomes zero. For $t \ll t^\ast$, the dynamics do not distinguish between the step-function tumbling rates and the homogeneous ones while for $t \gg t^\ast$, the inhomogeneity in the tumbling rates is felt  and the MSD leaves the plateau to relax towards the steady state. Generalizing this argument to arbitrary $u, g$, we expect that $t^*$ is determined by  
\be
{\bar x}_{homo}(t^*) \approx \min(u^{-1},g^{-1}) \label{tstarcrit}
\ee
Our simulation data in Fig. \ref{fig:u1} for $u=0$ and finite $g$ show that for $g \ll 1$,  in accordance with (\ref{tstarcrit}), the time scale $t^*$ is smaller compared to when $g \to \infty$ but it has a rather weak $g$-dependence for $g \gtrsim 1$.  
For inhomogeneous speed and tumbling rates, as shown in Fig.~\ref{fig:ug4}, we see a similar pattern in that the time scale $t^*$ decreases with decreasing $u$, as suggested by (\ref{tstarcrit}). 

For $\gamma_0 > \gamma_1$ where the tumbling rate never vanishes, instead of plateauing, the MSD varies in a nonmonotonic fashion as shown in Fig. \ref{fig:u0} for $u=0$ and various different $g$ values, and in Fig. \ref{fig:ug4} for different $u$ values and fixed $g$. As in the above discussion for $\gamma_0=\gamma_1$, the homogeneous limit considered in Appendix \ref{app_cnst} applies in this case also until a time $t^*$, and (\ref{varFl}) shows that the variance increases linearly with time. To find how long this diffusive growth continues, we again consider the $u=0, g \to \infty$ case for which our analytical calculations in Appendix \ref{app_msdn} indicate that the time $t^\ast$ where the MSD peaks is consistent with that  obtained from ${\bar x}_{homo}(t^\ast) = 0$. More explicitly, for $v_1=0$ in (\ref{meanFl}), we obtain 
\be 
t^\ast = \frac{x_0 \gamma_0}{\gamma_1 v_0} +\frac{(\Delta \gamma_0 + \gamma_1)}{2 \gamma_0 \gamma_1}. \label{eq:tstar}
\ee
which is $\approx  \frac{x_0 \gamma_0}{\gamma_1 v_0}$ for large $x_0$, as shown in Appendix \ref{app_msdn}. For general parameters, for the same reasons as for $\gamma_0=\gamma_1$, (\ref{tstarcrit}) is expected to hold, and the simulation data in Fig. \ref{fig:u0} and Fig. \ref{fig:ug4} are roughly consistent with it. 

The above discussion also gives an insight into how large the initial position be to observe a plateau or peak in the MSD. The intermediate regime, if it exists, starts when $t > (2 \gamma_0)^{-1}$ and lasts until time $t^*$ and therefore, if $(2 \gamma_0)^{-1} > t^*$,  the initial position is sufficiently small and the MSD will increase monotonically towards the stationary state otherwise a plateau or peak in the MSD dynamics will be observed. For example, for $\gamma_0=\gamma_1, u=0, g \to \infty$, when $x_0 < \dfrac{v_0}{2 \gamma_0}$, the plateau in the MSD is absent while for $x_0 \gg \dfrac{v_0}{2 \gamma_0}$, the MSD remains  stationary during $\dfrac{1}{2\gamma_0} \ll t \ll \dfrac{x_0}{v_0}$ and then relaxes towards the true stationary state. 

\subsection{Exponential relaxation at large times}
\label{exposec}

For $t > t^\ast$, as shown in Fig.~\ref{fig:u0} for uniform velocity and Fig.~\ref{fig:ug4} for inhomogeneous velocity, the MSD relaxes towards its steady state value $\sigma^2$ exponentially fast. For $u=0$, $g \to \infty$, in Appendices \ref{app_msd0} and \ref{app_msdn}, we calculate the relaxation time scale for $\gamma_0 = \gamma_1$ and $\gamma_0 \neq \gamma_1$, respectively, and find that 
\bsn
{|\sigma^2(x_0,t)-\sigma^2| \propto}
t^{-1/2} e^{-(\gamma_0-\sqrt{\gamma_0^2-\gamma_1^2}) t} ~&,~ $\gamma_0 \neq \gamma_1$ \label{expo1}\\
e^{-2 \gamma_0 t} ~&,~ $\gamma_0=\gamma_1$. \label{expo2}
\esn
In Fig.~\ref{fig:u0} that shows the MSD for $\gamma_0 \neq \gamma_1, u=0$ and various $g$, we verify (\ref{expo1}) for the relaxation time scale when $\gamma_0 \neq \gamma_1$ and $g \to \infty$; interestingly, we find that at long times, the  data for finite $g > 1$ is almost indistinguishable from the $g \to \infty$ limit. For $g=1/2$, although the time scale over which the system relaxes appears to be the same as in (\ref{expo1}), we expect the results to differ significantly as $g \to 0$ since no stationary state exists for $u=g=0$. Indeed our numerical simulations (data not shown) suggest that as $g \to 0$, the relaxation time scale in (\ref{expo1}) diverges as $g^{-1}$.

For  $u \neq 0$, we have not been able to calculate $\sigma^2(x_0, t)$ analytically. But our simulation results shown in  Fig. \ref{fig:ug4} indicate that the MSD relaxes exponentially fast, and a quantitative comparison between $u \to \infty$ and $u \gtrsim 1$ reveals that the relaxation time scale does not depend strongly on $u$.

\section{Summary and open questions}

In this work, we have considered a run-and-tumble particle in one dimension with spatially varying velocities and tumbling rates. Close to the origin, the spatial variation of velocities and tumbling rates is parameterized by $u$ and $g$, respectively, while at large distances from the origin, the spatial variation can be neglected and the rates saturate. For non-negative $u$ and $g$, we show that as long as at least one of these parameters is positive, the system reaches a stationary state. Even when one of these parameters is negative, a steady state can be reached, provided the other one has a large enough positive value. For example, a negative $u$ would mean that particle moves slower (faster) when it is heading towards (away from) the origin. This makes it easier for the particle to escape to infinity, unless $g$ has a large positive value which makes it tumble so frequently while moving away from the origin that it can not escape. Thus a large positive $g$ can counter the effect of negative $u$. Similarly, the case for negative $g$ and positive $u$ can be argued. A detailed discussion of the phase diagram of the general model is beyond the scope of this article and will be discussed elsewhere. Here, we mainly focused on non-negative $u$ and $g$ and calculated the exact position distribution in the steady state which is found to be unimodal or bimodal depending on the choice 
of the parameters.

The consequence of saturation in rates is seen most clearly in the dynamics of the particle away from the steady state. Our numerical simulations in this case show that in some parameter regimes, the MSD can pass through a plateau or be a nonmonotonic function of time before it finally relaxes to the stationary state; such unusual behavior is not known from earlier studies for any run-and-tumble motion. It would be very interesting to look for these features in other active systems and in experiments.

An insight into our numerical results mentioned above is obtained by our analytical calculations for the case of homogeneous speed ($u=0$) and step-function tumbling rates ($g \to \infty$). This limit has been previously studied in \cite{Singh:2020} only for initial position at the origin with an equal density of right- and left-movers. But here we have obtained an exact expression for the Green's function for an arbitrary initial position with arbitrary fraction of right- and left-movers present there, and calculated some properties of the MSD analytically. Although our calculations are restricted to the $g \to \infty$ case, we find that they quantitatively describe the simulation data even for finite $g \gtrsim 1$. In the same vein, it would be useful to obtain analytical results for the case of uniform tumbling rate ($g=0$) and step-function velocities ($u \to \infty$), which also maps onto the problem of run-and-tumble particle with homogeneous speed and tumbling rate in an external potential $V(x)=|x|$ as our simulations suggest a weak dependence on $u$; see \cite{sanjib_modx} for a discussion in this direction.

In this study as well as in much of the literature on active systems, the relaxation dynamics have been investigated, but the stationary state dynamics in general models, and especially those that exhibit a phase transition in the steady state, remain to be explored.  Moreover, here we have considered the dynamics of a single RTP but interesting collective behavior has been observed in systems of many interacting particles whose tumbling rates are modified because of the interaction \cite{Jose:2023, Tailleur:2008, Slowman:2016, Put:2019}. In chemotaxis and chemokinesis of a colony of bacterial cells like {\sl Myxococcus xanthus} or eukaryotic cells like {\sl Dictyostelium discoideum}, the cells modify their tumbling rates based on the concentration of the signaling molecules secreted by the cells themselves \cite{kearns1998mx}. In these systems, the local cell density often regulates the tumbling rate of an individual cell \cite{shi1996cell}, since each cell is able to sense the signaling molecules secreted by other cells in its neighborhood. In \cite{chou2003interplay}, a coarse-grained model was introduced to study the behavior of such systems at the level of a single cell with run-and-tumble motion, whose tumbling rate is governed by its own chemical trail. It would be of interest to extend this study to a colony of cells which interact among themselves. Not only for microorganisms, run-and-tumble description has been used even for modeling formation and movement of animal herds \cite{eftimie2007complex}. Moving animals can change their direction of motion based on the information they receive from other animals in the group. The exchange of information between a pair depends on the distance between them, and also has certain directional aspects. Based on whether the pair is approaching each other or going away from each other, and whether the information is coming from behind or from some position ahead of the recipient, the information is processed differently. In \cite{eftimie2007complex}, interacting run-and-tumble particles were used to model such situations and different types of interactions were studied based on the above considerations, which successfully explained formation of complex spatial patterns in animal herds. It will be of interest to study the relaxation properties, and steady state dynamics of such interacting run-and-tumble particles.

\section{Acknowledgements}

KJ thanks Manasvi Gautam and Lakshita Jindal for many useful discussions during the early stages of this work. SC acknowledges support from Anusandhan National Research Foundation (ANRF), India (Grant No: CRG/2023/000159).

\clearpage

\appendix
\setcounter{equation}{0}
\renewcommand{\thesection}{\Alph{section}}
\numberwithin{equation}{section}


\section{Stationary state distribution}
\label{app_ss}

For completeness, here we briefly review the derivation of the stationary state distribution \cite{Schnitzer:1993,Monthus:2021}. 
In the stationary state, the left-hand side (LHS) of (\ref{Pp}) and (\ref{Pm}) obeyed by the distributions
$P_\pm(x,t)$ vanishes, 
\bea
 -(v_+(x) P_+(x))'+ \gamma_-(x) P_-(x) -\gamma_+(x) P_+(x) &=& 0 \label{app_Pp} \\
 (v_-(x) P_-(x))'+ \gamma_+(x) P_+(x) -\gamma_-(x) P_-(x) &=&0 \label{app_Pm}
 \eea
 As the total current $J(x)=J_+(x)+J_-(x)$ with $J_\pm(x)=\pm v_\pm(x) P_\pm(x)$ is uniform in the stationary state and must vanish at $x \to \pm \infty$, we also have 
\be
J_+(x)+J_-(x)=v_+(x) P_+(x)-v_-(x) P_-(x)=0 ~~\forall~~ x \label{app_J}
\ee
Using (\ref{app_J}) in  (\ref{app_Pp}) and  (\ref{app_Pm}), we immediately obtain
 \bea
 P_+(x) &=& \frac{v_+(0) P_+(0)}{v_+(x)} e^{\int_0^x \left( \frac{\gamma_-(y)}{v_-(y)}- \frac{\gamma_+(y)}{v_+(y)} \right) dy} \label{app_Pp2} \\
 P_-(x)&=& \frac{v_-(0) P_-(0)}{v_-(x)} e^{\int_0^x \left( \frac{\gamma_-(y)}{v_-(y)}- \frac{\gamma_+(y)}{v_+(y)} \right) dy} \label{app_Pm2}
 \eea
 provided $v_\pm(x) \neq 0$ for $-\infty < x < \infty$ (for a discussion of stationary state when the speeds vanish, see \cite{Dhar:2019}). Except for $u \to \infty, v_0=v_1$ where $v_\pm(x)=v_0 [1 \mp \textrm{sgn}(x)]$, as (\ref{vchoice}) satisfies this condition, the total stationary state probability distribution, $P(x)=P_+(x)+P_-(x)$ can be written as  
 \bea
 P(x) &=& \left[\frac{J_+(0)}{v_+(x)} -\frac{J_-(0)}{v_-(x)} \right]e^{\int_0^x \left( \frac{\gamma_-(y)}{v_-(y)}- \frac{\gamma_+(y)}{v_+(y)} \right) dy} \\
&=&  J_+(0) \left[\frac{1}{v_+(x)} +\frac{1}{v_-(x)} \right]e^{\int_0^x \left( \frac{\gamma_-(y)}{v_-(y)}- \frac{\gamma_+(y)}{v_+(y)} \right) dy}
 \eea
Due to particle conservation, $P(x)$ must be normalizable and therefore, the stationary state exists if  
 \bea
{\cal N}=J_+(0) \int_{-\infty}^\infty dx  \left[\frac{1}{v_+(x)} +\frac{1}{v_-(x)} \right]e^{\int_0^x \left( \frac{\gamma_-(y)}{v_-(y)}- \frac{\gamma_+(y)}{v_+(y)} \right) dy} 
\eea
is finite. We also note that for the rates given by (\ref{vchoice}) and (\ref{gchoice}), $v_\pm(-x)=v_\mp(x), \gamma_\pm(-x)=\gamma_\mp(x)$ using which in (\ref{app_Pp2}) and (\ref{app_Pm2}), it immediately follows that $P_+(-x)=P_-(x)$ and therefore, $\int_{-\infty}^\infty dx' P_\pm(x')=1/2$. 

For the special case where $u \to \infty, v_0=v_1, g \geq 0$, as $v_+(x)=0, x > 0$, it follows from (\ref{app_J}) and (\ref{app_Pp}) that $P_\mp (x)=0, x > 0$; similarly, as $v_-(x)=0, x < 0$, we obtain $P_\pm(x)=0, x <0$. Taken together, these imply that $P(x)=\delta(x)$ in this case.

\section{Uniform asymmetric rates with asymmetric initial condition}
\label{app_cnst}

To understand the dynamics of the mean-squared displacement, we first note that for arbitrary rates, on taking the Laplace transform ${\tilde P}_\pm(x,s)=\int_0^\infty P_\pm(x,t) e^{-s t} dt$ of (\ref{Pp}) and (\ref{Pm}) on both sides, we have 
\bea
s {\tilde P}_+(x,s) - P_+(x,0) &=& -[v_+(x) {\tilde P}_+(x,s)]'+ \gamma_-(x) {\tilde P}_-(x,s) -\gamma_+(x) {\tilde P}_+(x,s) \label{LT1} \\
s {\tilde P}_-(x,s) - P_-(x,0) &=& [v_-(x) {\tilde P}_-(x,s)]'+ \gamma_+(x) {\tilde P}_+(x,s) -\gamma_-(x) {\tilde P}_-(x,s)  \label{LT2}
\eea
where, $P_\pm(x,0)$ are the initial distributions given by (\ref{initialcond}). 

As explained in Sec.~\ref{sec:dyn}, the short time dynamics can be captured if one assumes that the rates are given by  (\ref{univ}) and (\ref{unig}), $v_{\pm}(x)=v_0\mp v_1, \gamma_{\pm}(x)=\gamma_0\pm \gamma_1$ for all $x$. For general homogeneous rates (that is, $\gamma_\pm(x)=\gamma_\pm, v_\pm(x)=v_\pm$), on eliminating ${\tilde P}_-(x,s)$ from (\ref{LT1}) and substituting it in (\ref{LT2}), we obtain a second order, inhomogeneous differential equation for ${\tilde P}_+(x,s)$; in a similar manner, one can find the equation obeyed by ${\tilde P}_-(x,s)$. Together these lead to an equation for ${\tilde P}_{homo}(x,s)={\tilde P}_+(x,s)+{\tilde P}_-(x,s)$ which is given by 
\bea
{\tilde P}_{homo}''(x,s)+\frac{v_- (\gamma_++s)-v_+(\gamma_-+s)}{v_- v_+} {\tilde P}'_{homo}(x,s) -\frac{s (\gamma_-+\gamma_++s)}{v_- v_+}  {\tilde P}_{homo}(x,s)={\tilde H}_{homo}(x,s) \label{cnstPpeqn}
\eea
where the inhomogeneous term,
\be
{\tilde H}_{homo}(x,s; x_0) =\frac{[v_- P'_+(x,0)-v_+ P'_-(x,0)]-[s+\gamma_-+\gamma_+] P(x,0)}{v_- v_+} \label{homoin}
\ee
and we have added the subscript in the above expressions to emphasize that these hold for the homogeneous rates  (\ref{univ}) and (\ref{unig}). 
 
To solve the inhomogeneous equation (\ref{cnstPpeqn}), we first find the Green's function for which $L_{x,s} {\tilde G}_{homo}(x,s;a)=\delta(x-a)$ where $L_{x,s}$ is the differential operator on the LHS of (\ref{cnstPpeqn}).  
If the homogeneous equation $L_{x,s} {\tilde G}_{homo}=0$ has two linearly independent solutions denoted by $y_1(x,s)$ and $y_2(x,s)$ where $y_1(x,s) \to 0$ as $x \to -\infty$ and $y_2(x,s) \to 0$ as $x \to +\infty$, we obtain
\bsn
{ {\tilde G}_{homo}(x,s; a)=\label{Ghomofull}}
c_1(a,s) y_1(x,s) ~&,~ $x < a$ \label{Gdefn1}\\
c_2(a,s) y_2(x,s) ~&,~ $x > a$ \label{Gdefn2}
\esn
where the constants $c_1, c_2$ are obtained by demanding the continuity of solutions at $x=a$ and a finite discontinuity in the first derivative of the Green's function at $x=a$. We thus obtain $c_1(a,s)=\frac{y_2(a,s)}{W(a,s)}, c_2(a,s)=\frac{y_1(a,s)}{W(a,s)}$ where $y_1=e^{\alpha_+ x}, y_2=e^{\alpha_- x}$ and the Wronskian, $W(a,s)=(\alpha_- - \alpha_+) e^{(\alpha_- + \alpha_+) x_0}$, and $\alpha_+ > 0$ and $\alpha_- < 0$ are the solutions of the following quadratic equation, 
\be
\alpha^2+\frac{v_- (\gamma_++s)-v_+(\gamma_-+s)}{v_- v_+} \alpha -\frac{s (\gamma_-+\gamma_++s)}{v_- v_+}=0 \label{root}
\ee
For later reference, we note that for $v_\pm=v_0, \gamma_\pm=\gamma_0 \pm \gamma_1$,  the inverse Laplace transform of ${\tilde G}_{homo}(x,s;a)$ is given by
\be
G_{homo}(x,t; a)=-\frac{v_0 e^{\frac{-\gamma_1(x-a)}{v_0}}}{2} e^{-\gamma_0 t} I_0 \left[\sqrt{\gamma_0^2-\gamma_1^2}  \sqrt{t^2-\left(\frac{x-a}{v_0} \right)^2} \right] \Theta (v_0 t-|x-a|) \label{homoGv0} 
\ee 
where $I_n(z)$ is the modified Bessel function of the first kind. 

For general homogeneous rates, using  (\ref{initialcond}) in (\ref{homoin}), and (\ref{Ghomofull}), we find that
\bsn
{{\tilde P}_{homo}(x,s)=\int_{-\infty}^\infty  da {\tilde G}_{homo}(x,s; a) {\tilde H}_{homo}(x,s; a)=}
\frac{s+(\gamma_-+\gamma_+)+(v_+ a_- -v_- a_+) \alpha_+ }{v_- v_+ (\alpha_+ - \alpha_-)} \frac{y_1(x,s)}{y_1(x_0,s)}   ,& $x < x_0$ \\
\frac{s +(\gamma_- +\gamma_+)+(v_+ a_- -v_- a_+) \alpha_-}{v_- v_+  (\alpha_+ - \alpha_-)} \frac{y_2(x,s)}{y_2(x_0,s)} ,& $x > x_0$ 
\esn
For the rates (\ref{univ}) and (\ref{unig}), after inverting the Laplace transform,  we obtain the following exact expressions for the mean and variance in the position of the particle: 
\bea
{\bar x}_{homo}(t)
   &=& x_0-\frac{t (\gamma_0 v_1+\gamma_1 v_0)}{\gamma_0}+\frac{v_0 (\Delta \gamma_0+\gamma_1)}{2 \gamma_0^2} -\frac{v_0 e^{-2 \gamma_0 t} (\Delta \gamma_0+\gamma_1)}{2 \gamma_0^2} \label{cstmean}\\
\frac{\sigma_{homo}^2(t)}{v_0^2} &=& \frac{(\gamma_0^2-\gamma_1^2) t}{\gamma_0^3}+\frac{5 \gamma_1^2-2 \gamma_0^2-\Delta^2 \gamma_0^2+2 \Delta \gamma_0 \gamma_1}{4 \gamma_0^4}\nn \\
   &+&e^{-2 \gamma_0 t} \left[\frac{\gamma_0 \left(\left(\Delta^2+1\right) \gamma_0-4 \gamma_1 t (\Delta
   \gamma_0+\gamma_1)\right)-2 \gamma_1^2}{2 \gamma_0^4} \right] \nn \\
   &-&e^{-4 \gamma_0 t} \left[\frac{(\Delta \gamma_0+\gamma_1)^2}{4 \gamma_0^4} \right] \label{cstvar}
   \eea
   where $\Delta=a_+ - a_-$. Note that the above expression for variance is independent of $v_1$. 
%

\section{Uniform speed, step-function tumbling rates: exact Green's function}
\label{app_Green}

Here we calculate the exact Green's function when $u=0, g \to \infty$ in (\ref{vchoice}) and (\ref{gchoice}) so that $v_\pm(x)=v_0, \gamma_\pm(x)=\gamma_0+\gamma_1 \textrm{sgn}(x)=\gamma_0+\gamma_1 [2 \Theta(x)-1]$. 
The Laplace transform of $P_\pm(x,t)$ obey (\ref{LT1}) and (\ref{LT2}), and for uniform speed but  arbitrary, inhomogeneous tumbling rates, it is convenient to work with ${\tilde Q}(x,s)={\tilde P}_+(x,s)-{\tilde P}_-(x,s)$ and ${\tilde P}(x,s)={\tilde P}_+(x,s)+{\tilde P}_-(x,s)$. From (\ref{LT1}) and (\ref{LT2}), we then obtain
\bea
s  {\tilde P}(x,s)-P(x,0) &=& -v_0 {\tilde Q}'(x,s) \label{Qtildeprime}\\
s  {\tilde Q}(x,s)- Q(x,0) &=& -v_0 {\tilde P}'(x,s)-2 \gamma_0 {\tilde Q}(x,s)-2 \gamma_1 (2 \Theta(x)-1) {\tilde P}(x,s) \label{Ptildeprime}
\eea
Eliminating ${\tilde Q}(x,s)$ from (\ref{Ptildeprime}) and plugging it in (\ref{Qtildeprime}), we find that ${\tilde P}(x,s)$ obeys a second order ordinary differential equation, 
\bea
 {\tilde P}^{''}(x,s)+\frac{2}{v_0} [2 \gamma_1 \delta(x)  {\tilde P}(x,s)+\gamma_1 (2 \Theta(x)-1) {\tilde P}^{'}(x,s)]  -\frac{s (2 \gamma_0+s)}{v_0^2}  {\tilde P}(x,s) &=& {\tilde H}(x,s; x_0) \label{Ptildeprime2}
\eea
where, 
\be
{\tilde H}(x,s; x_0)=\frac{v_0 Q'(x,0)-(2 \gamma_0+s) P(x,0)}{v_0^2}=\frac{\Delta}{v_0} \delta'(x-x_0)-\frac{2 \gamma_0+s}{v_0^2} \delta(x-x_0)  \label{Ptildeprime3}
\ee
on using (\ref{initialcond})  and the asymmetry parameter, $\Delta = a_+-a_-$. Throughout the article, we assume that $x_0 \geq 0$ but from (\ref{Ptildeprime2}) and (\ref{Ptildeprime3}), we also note that
\be
{\tilde P}(x,s; x_0, \Delta)={\tilde P}(-x,s; -x_0, -\Delta)
\ee
Note that the coefficients of (\ref{Ptildeprime2}) are not smooth functions of $x$. Using the Green's function ${\tilde G}(x,s; a)$ that obeys (\ref{Greenv1z}) below, we then have  
\bea
\tilde P(x, s) &=&\int_{-\infty}^\infty {\tilde G}(x,s; a) {\tilde H}(x, s; a) da \\
&=& -\int_{-\infty}^\infty \left[\frac{\Delta}{v_0} \frac{d {\tilde G}(x,s;a)}{da}+\frac{2 \gamma_0+s}{v_0^2} {\tilde G}(x,s;a) \right] \delta(a-x_0) da \label{Pxsint}
\eea
where we have carried out an integration by parts to arrive at the last expression. 

For the inhomogeneous equation (\ref{Ptildeprime2}), the Green's function obeys
 \bea
 {\tilde G}^{''}(x,s)+\frac{2}{v_0} [2 \gamma_1 \delta(x)  {\tilde G}(x,s)+\gamma_1 (2 \Theta(x)-1) {\tilde G}^{'}(x,s)]  -\frac{s (2 \gamma_0+s)}{v_0^2}  {\tilde G}(x,s) &=&\delta(x-a) \label{Greenv1z}
\eea
where we have suppressed the $a$-dependence in ${\tilde G}(x,s; a)$ for brevity. 
Assuming continuity of the homogeneous solutions of (\ref{Greenv1z}) at $x=a$ and on integrating (\ref{Greenv1z}) in the neighborhood of $a$, we get the usual condition for the discontinuity in its first derivative, 
\be
\text{Lim}_{\epsilon \to 0} {\tilde G}^{'}(a+\epsilon, s)-{\tilde G}^{'}(a-\epsilon, s)=1, a \neq 0 \label{discontn0}
\ee
provided $a$ is nonzero. To deal with the $\delta$-function on the LHS of (\ref{Greenv1z}), we integrate it over $x$ on both sides in the neighborhood of origin and obtain
\bea
\text{Lim}_{\epsilon \to 0} {\tilde G}^{'}(\epsilon, s)-{\tilde G}^{'}(-\epsilon, s)+\frac{4 \gamma_1 }{v_0} {\tilde G}(0, s) =\delta_{a,0} \label{discont0}
\eea
where we have assumed that the Green's function is continuous but not differentiable at the origin. 
For $x \neq 0$, the solution of the homogeneous equation (\ref{Greenv1z}) is given by
\bsn
{{\tilde G}(x, s)=}
a_1 e^{\alpha_+ x}+a_2 e^{\alpha_- x} ~&,~ $x > 0$ \\
b_1 e^{-\alpha_- x}+b_2 e^{-\alpha_+ x} ~&,~ $x < 0$
\esn
where $\alpha_+ > 0,  \alpha_- <  0$ are solutions of $\alpha^2+\frac{2 \gamma_1}{v_0} \alpha -\frac{s (2 \gamma_0+s)}{v_0^2}=0$. 
For $x_0 > 0$ (as assumed in this article), as (\ref{Pxsint}) is nonzero for $a > 0$, we assume this condition for the discussion below. For $a > 0$, on using that the solution ${\tilde P}(x, s)$ and therefore, ${\tilde G}(x, s)$ must vanish at $x \to \pm \infty$, we get
\bsn
{{\tilde G}(x, s)=\label{Gpeqn}}
a_2 e^{\alpha_- x}  ~&,~ $x > a$ \\
c_1 e^{\alpha_+ x}+c_2 e^{\alpha_- x} ~&,~ $0 < x < a$\\
b_1 e^{-\alpha_- x}   ~&,~ $x < 0, x < a$
\esn
Using continuity of ${\tilde G}(x, s)$ at $x=0$ and $x=a$, and (\ref{discontn0}) and (\ref{discont0}), all four unknowns in the above equations can be determined, and are given by
\bea
\alpha_\pm &=& \frac{-\gamma_1 \pm \sqrt{(\gamma_0^2-\gamma_1^2) (w+2) w}}{v_0} \\
a_2 &=&  \frac{e^{-a \alpha_-}}{\alpha_--\alpha_+} +\frac{\gamma_1 e^{-a \alpha_+}}{v_0
   (\alpha_--\alpha_+) \alpha_+}\\
c_1 &=&   \frac{e^{-a \alpha_+}}{\alpha_--\alpha_+} \\
c_2 &=& \frac{ \gamma_1 e^{-a \alpha_+}}{v_0
   (\alpha_--\alpha_+) \alpha_+} \\
b_1 &=&  -\frac{e^{-a \alpha_+}}{2 \alpha_+} 
\eea
where $w = \frac{s+\gamma_0-\sqrt{\gamma_0^2-\gamma_1^2}}{\sqrt{\gamma_0^2-\gamma_1^2}}$ as in \cite{Singh:2020}.

To find the Green's function in real space, we need to invert the Laplace transforms. Using (29.2.14) and (29.3.91) of \cite{Abramowitz:1964}, we immediately obtain
\bea
L^{-1}[c_1 e^{\alpha_+ x}] {=}-\frac{v_0}{2} e^{\frac{\gamma_1}{v_0} (a-x)} e^{-\gamma_0 t} I_0 \left[\sqrt{\gamma_0^2-\gamma_1^2}  \sqrt{t^2-\left(\frac{a-x}{v_0} \right)^2} \right] \Theta \left[t-\frac{a-x}{v_0} \right] \label{LTi} 
\eea
where $L^{-1} f(s)$ denotes the inverse Laplace transform of $f(s)$ and $I_n(z)$ is the modified Bessel function of first kind. Using the derivative of (29.3.91) of \cite{Abramowitz:1964} with respect to $k$, we first note that
\bea
&&L^{-1}[\frac{d}{da} (b_1 e^{-\alpha_- x})] {=}  \frac{e^{\frac{\gamma_1 (x+a)}{v_0}}}{2} e^{-\gamma_0 t}
\delta \left(t-\frac{a-x}{v_0} \right)+\frac{\frac{a-x}{v_0} \Theta (t-\frac{a-x}{v_0}) I_1\left(\sqrt{\gamma_0^2-\gamma_1^2}\sqrt{t^2-(\frac{a-x}{ v_0})^2}\right)}{\sqrt{t^2-(\frac{a-x}{v_0})^2}}
\eea
We fix the constant of integration arising due to the integral over $a$ by demanding that $b_1 e^{-\alpha_- x}$ and its inverse Laplace transform vanish for $a \to \infty$, and obtain
\bea
L^{-1}[b_1 e^{-\alpha_- x}]&{=}& -\Theta(x+v_0 t-a) \frac{v_0 e^{-\gamma_0 t}}{2} e^{\frac{\gamma_1 (x+a)}{v_0}} I_0\left[\sqrt{\gamma_0^2-\gamma_1^2}\sqrt{t^2-\left(\frac{a-x}{v_0}\right)^2}\right]\nn \\
&-& \Theta(x+v_0 t-a) \frac{v_0 e^{-\gamma_0 t}}{2} \int_a^{x+v_0 t} da' \frac{\gamma_1}{v_0} e^{\frac{\gamma_1 (x+a')}{v_0}} I_0\left[\sqrt{\gamma_0^2-\gamma_1^2}\sqrt{t^2-\left(\frac{a'-x}{v_0}\right)^2}\right] \label{LTii}
\eea
after an integration by parts. Proceeding in a fashion similar to above, we also obtain
\bea
L^{-1}[a_2 e^{\alpha_- x}] &=& -\frac{v_0}{2} e^{\frac{\gamma_1}{v_0} (a-x)} e^{-\gamma_0 t} I_0 \left[\sqrt{\gamma_0^2-\gamma_1^2}  \sqrt{t^2-\left(\frac{a-x}{v_0} \right)^2} \right] \Theta (v_0 t-x+a)\nn \\
&-& \Theta(v_0 t-x -a) \int_a^{v_0 t-x} da'\frac{\gamma_1 e^{-\gamma_0 t} e^{\frac{\gamma_1 (a'-x)}{v_0}}}{2} I_0\left[\sqrt{\gamma_0^2-\gamma_1^2} \sqrt{t^2-\left(\frac{a'+x}{v_0} \right)^2}\right] \label{LTiii}\\
L^{-1}[c_2 e^{\alpha_- x}] &=&  -\Theta(v_0 t-x -a) \int_a^{v_0 t-x} da'\frac{\gamma_1 e^{-\gamma_0 t} e^{\frac{\gamma_1 (a'-x)}{v_0}}}{2} I_0\left[\sqrt{\gamma_0^2-\gamma_1^2}\sqrt{t^2-\left(\frac{a'+x}{v_0} \right)^2}\right]  \label{LTiv}
\eea

Using (\ref{LTi})-(\ref{LTiv}) in (\ref{Gpeqn}), we finally arrive at the following {exact} expression for the Green's function for $a > 0$ and arbitrary $x$:
\bea
&&G(x,t; a) {=} -\frac{v_0}{2} e^{\frac{\gamma_1}{v_0} (a-|x|)} e^{-\gamma_0 t} I_0 \left[\sqrt{\gamma_0^2-\gamma_1^2}  \sqrt{t^2-\left(\frac{a-x}{v_0}\right)^2} \right] \Theta(v_0 t-|x-a|)\nn \\
&-&\Theta(v_0 t-|x| -a) \int_a^{v_0 t-|x|} da'\frac{\gamma_1 e^{-\gamma_0 t} e^{\frac{\gamma_1 (a'-|x|)}{v_0}}}{2} I_0\left[\sqrt{\gamma_0^2-\gamma_1^2} \sqrt{t^2-\left(\frac{a'+|x|}{v_0}\right)^2} \right] \label{finalG}
\eea
Note that in the above expression, the first term on the RHS clicks at all times but the second term is nonzero only when $t > \frac{x_0}{v_0}$. As we will see in Appendix~\ref{app_msd0} and Appendix~\ref{app_msdn},  the first term in the above Green's function captures the dynamics when the particle experiences homogeneous rates and its mean position is $> 0$ (compare the first term above  for positive $x$ with (\ref{homoGv0})), while  both terms control the long-time relaxation dynamics where the first term vanishes and second term approaches the stationary state as $t \to \infty$. 
Taking the inverse Laplace transform of (\ref{Pxsint}) on both sides, we find that the exact probability distribution $P(x,t; x_0)$ is given by
\bea
P(x, t; x_0) = -\frac{\Delta}{v_0} \frac{\partial G(x,t;x_0)}{\partial x_0}-\frac{2 \gamma_0}{v_0^2} G(x,t;x_0) -\frac{1}{v_0^2} \frac{\partial G(x,t;x_0)}{\partial t} \label{ILTdistform2} 
\eea
where we have used (29.2.4) of \cite{Abramowitz:1964} and that $G(x,t;x_0)=0$ due to (\ref{finalG}). In the stationary state where the probability distribution is independent of $t$ and $x_0$, on using (\ref{v1zerostep}) in (\ref{ILTdistform2}), we obtain the stationary state Green's function to be
\be
G(x)=-\frac{v_0 e^{-\frac{2 \gamma_1 |x|}{v_0}}}{2} \label{stepGss}
\ee
In Appendix~\ref{app_msd0} and Appendix~\ref{app_msdn}, it is verified that (\ref{finalG}) indeed reduces to the above equation in the steady state. We have also checked that (\ref{finalG}) and (\ref{ILTdistform2}) reproduce the result for $x_0=0$ obtained in \cite{Singh:2020}. 

\section{Uniform speed, step-function tumbling rates: $\gamma_0=\gamma_1$}
\label{app_msd0}

For $\gamma_0=\gamma_1$, the Green's function (\ref{finalG}) simplifies to
\bea
G(x,t; a) &=&-\frac{v_0}{2} e^{\frac{\gamma_0}{v_0} (a-|x|)} e^{-\gamma_0 t}  [\Theta(v_0 t-|x-a|)-\Theta(v_0 t-|x| -a)] \nn \\
 &+& \Theta(v_0 t-|x| -a) G(x) \label{g0g1Gfn}
 \eea
 which approaches the stationary state Green's function $G(x)$ given by (\ref{stepGss}) when $t \to \infty$.

 Using the above expression in (\ref{ILTdistform2}), the exact position probability distribution and its cumulants can be obtained; in particular, the mean and variance of the position of the particle are given by 
 \bsn
{{\bar x}(t)=\label{meaneq}}
 x_0-v_0 t+\frac{(1+\Delta) v_0 (1-e^{-2 \gamma_0 t})}{2 \gamma_0}~&,~ $t < \frac{x_0}{v_0}$ \\
   r_0(t) e^{-2 \gamma_0 t} ~&,~ $t > \frac{x_0}{v_0}$
 \esn
 and
 \bsn
{\frac{\sigma^2(x_0,t)}{v_0^2}=\label{msdeq}}
\frac{(3-\Delta)(1+\Delta)}{4
   \gamma_0^2}+\frac{(\Delta+1)  (\Delta-1-4 \gamma_0 t) e^{-2 \gamma_0 t}}{2 \gamma_0^2} - \frac{(\Delta+1)^2 e^{-4 \gamma_0 t}}{4 \gamma_0^2} ~&,~ $t < \frac{x_0}{v_0}$ \\
 \frac{1}{2 \gamma_0^2}+r_1(t) e^{-2 \gamma_0 t}  
   - r_2(t) e^{-4 \gamma_0 t} ~&,~ $t > \frac{x_0}{v_0}$
   \esn
where,
\bea
r_0(t) &=& \frac{e^{\frac{2 \gamma_0 x_0}{v_0}} (v_0 (2 \Delta \gamma_0
   t+\Delta+1)-2 \Delta \gamma_0 x_0)-(\Delta+1) v_0}{2
   \gamma_0} \\
   r_1(t) &=&  \frac{e^{\frac{2 \gamma_0 x_0}{v_0}} (2 \Delta v_0 (\gamma_0
   t+1)-4 \Delta \gamma_0 x_0+v_0)-2 (\Delta+1) (\gamma_0 t
   v_0+\gamma_0 x_0+v_0)}{2 \gamma_0^2 v_0} \\
   r_2(t) &=& \frac{\left(e^{\frac{2 \gamma_0 x_0}{v_0}} (2 \Delta \gamma_0
   x_0-v_0 (2 \Delta \gamma_0 t+\Delta+1))+(\Delta+1)
   v_0\right)^2}{4 \gamma_0^2 v_0^2}
  \eea
  On setting $v_1=0, \gamma_0=\gamma_1$ in (\ref{cstmean}) and (\ref{cstvar}), we find that they match (\ref{meaneq}) and (\ref{msdeq}), respectively,  when $t < \frac{x_0}{v_0}$ which shows that the short time dynamics are exactly described by those of the RTP with homogeneous rates.      
  
 The above expressions show that the dynamics depend on whether time is above or below $\frac{x_0}{v_0}$ and involve terms that decay exponentially fast with the slowest time scale being $(2 \gamma_0)^{-1}$.  
  Below we give approximate results for the mean position and mean-squared displacement depending on these time scales. For $x_0 > \frac{v_0}{2 \gamma_0}$, we find that 
\bsn
{{\bar x}(t) \approx \label{app_mean01}} 
x_0+\Delta v_0 t~&,~$t \ll \frac{1}{2 \gamma_0} < \frac{x_0}{v_0}$ \\
x_0+\frac{(1+\Delta) v_0}{2 \gamma_0}-t
   v_0  ~&,~$\frac{1}{2 \gamma_0} \ll t < \frac{x_0}{v_0}$ \label{meaneqinter} \\
r_0(t) e^{-2 \gamma_0 t}  ~&,~$t > \frac{x_0}{v_0}$
\esn
and
\bsn
{\sigma^2(x_0,t) \approx \label{app_msd01}}
(1-\Delta^2) v_0^2 t^2 ~&,~$t \ll \frac{1}{2 \gamma_0} < \frac{x_0}{v_0}$ \\
\frac{(3-\Delta) (\Delta+1) v_0^2}{4 \gamma_0^2} ~&,~$\frac{1}{2 \gamma_0} \ll t < \frac{x_0}{v_0}$ \\
\sigma^2+r_1(t) e^{-2 \gamma_0 t} ~&,~$t > \frac{x_0}{v_0}$ 
 \esn
 where $\sigma^2=\frac{v_0^2}{2 \gamma_0^2}$ is the stationary state variance as shown in Sec.~\ref{tumble}.  
  The above expressions thus show that the variance initially increases superdiffusively and becomes a constant at intermediate times, but for time $t > t^*=\frac{x_0}{v_0}$, these dynamics are followed by an exponential relaxation at rate $2 \gamma_0$ towards the stationary state. The mean displacement, however, decreases monotonically towards zero - linearly at short times and exponentially fast at longer times.  In contrast, for $x_0 < \frac{v_0}{2 \gamma_0}$, both mean and variance approach the steady state in a monotonic fashion, 
 \bsn
{{\bar x}(t) \approx} 
x_0+\Delta v_0 t~&,~$t \ll \frac{1}{2 \gamma_0} < \frac{x_0}{v_0}$ \\
   v_0  ~&,~$\frac{1}{2 \gamma_0} \ll t < \frac{x_0}{v_0}$ \\
r_0(t) e^{-2 \gamma_0 t}  ~&,~$t > \frac{x_0}{v_0}$
\esn
and 
\bsn
{\sigma^2(x_0,t) \approx \label{app_msd02}}
(1-\Delta^2) v_0^2 t^2 ~&,~$t \ll\frac{x_0}{v_0} < \frac{1}{2 \gamma_0} $ \\
\sigma^2+r_1(t) e^{-2 \gamma_0 t} ~&,~$t > \frac{1}{2 \gamma_0}$ 
 \esn

Thus, in either case, the stationary state is approached over a time scale $(2 \gamma_0)^{-1}$ but the relaxation regime starts at $\max(\frac{x_0}{v_0},\frac{1}{2 \gamma_0})$. To understand the meaning of $t^*=\frac{x_0}{v_0}$, from (\ref{meaneqinter}), we note that for $x_0 \gg \frac{v_0}{2 \gamma_0}$, the mean position ${\bar x}(t) \approx x_0 -t v_0$ is at the origin at time $t^*$ and thus the particle is sensitive to the inhomogeneous tumbling rate beyond this time scale.  This discussion also shows that if this time is much smaller than the typical time for tumbling to occur, that is, $t^* \ll (2 \gamma_0)^{-1}$, the variance will show monotonic dynamics. 

We also mention that in the special case when initially there is only the left-mover  ($\Delta=-1$), (\ref{app_msd01}) and (\ref{app_msd02}) show that the variance remains zero (as expected) until $\max(\frac{x_0}{v_0},\frac{1}{2 \gamma_0})$ and then increases monotonically towards $\sigma^2$.

\section{Uniform speed, step-function tumbling rates: $\gamma_0 \neq \gamma_1$}
\label{app_msdn}

For $\gamma_0 \neq \gamma_1$, although using the exact Green's function (\ref{finalG}) in (\ref{ILTdistform2}), the exact probability distribution can be obtained, it does not seem possible to obtain the cumulants of mean displacement exactly, and therefore, below we develop some approximations for them. 

At large times, using the asymptotic expansion of $I_0(z) \stackrel{z \to \infty}{\sim} \frac{e^z}{\sqrt{2 \pi z}}$ in (\ref{finalG}), we obtain
\bea
G(x,t; a)& \approx & -\frac{v_0}{2} e^{\frac{\gamma_1}{v_0} (a-|x|)} e^{-\gamma_0 t} \frac{e^{t\sqrt{\gamma_0^2-\gamma_1^2}  \sqrt{1-(\frac{a-x}{v_0 t})^2}}}{\sqrt{2 \pi t \sqrt{\gamma_0^2-\gamma_1^2}  \sqrt{1-(\frac{a-x}{v_0 t})^2}}} \Theta(v_0 t-|x-a|)\nn \\
&-&\Theta(v_0 t-|x| -a) v_0 t \int_{\frac{a+|x|}{v_0 t}}^1 dw \frac{\gamma_1 e^{-\gamma_0 t} e^{\gamma_1 t w} e^{\frac{-2 \gamma_1 |x|}{v_0}}}{2} \frac{e^{t \sqrt{\gamma_0^2-\gamma_1^2}  \sqrt{1-w^2}}}{\sqrt{2 \pi t \sqrt{\gamma_0^2-\gamma_1^2}  \sqrt{1-w^2} }} \label{latetimeG}
\eea
In the above equation, as the second term on the RHS begins to contribute when $t > \frac{x_0}{v_0}$, we focus on the first term at short times. We first note that the first term is a nonmonotonic function of $x$ with a maximum at $x_*(a)=a - \frac{\gamma_1 v_0 t}{\gamma_0}$. Since for sufficiently large and positive $x_0$, the particle is likely to be at $x > 0$ at short times, we approximate the Green's function $G(x,t;x_0)$ for $t < \frac{x_0 \gamma_0}{v_0 \gamma_1}$ by a Gaussian centred about $x_*(x_0)$, and find the mean and variance in the particle's position using  (\ref{ILTdistform2}). We obtain
\bea
\bar{x}(t) &\approx& x_0+\frac{v_0 (\Delta \gamma_0-\gamma_1)}{2 \gamma_0^2}-\frac{t \gamma_1 v_0}{\gamma_0}  \label{meang0g1G}\\
\sigma^2(x_0,t) &\approx& \frac{(\gamma_0^2-\gamma_1^2) v_0^2 t}{\gamma_0^3} \label{varg0g1G}
\eea
For the mean displacement, we observe that the sign of third term in  (\ref{meang0g1G}) and that in (\ref{meanFl}) for homogeneous rates and $v_1=0$ do not match, but this appears to be a result of the Gaussian approximation as (\ref{homoGv0}) also yields (\ref{meang0g1G}) under this approximation; 
however, for the variance, (\ref{varg0g1G}) above and (\ref{varFl})  for $v_1=0$ match each other. These results are thus consistent with our expectation that  the particle with constant speed and step-function tumbling rates behaves like the RTP with homogeneous rates until the mean position of the particle becomes zero.

We now turn to a discussion of the late-time behavior to verify that the stationary state is obtained and to find the relaxation time scale. For this purpose, we need to consider the second term on the RHS of (\ref{latetimeG}); however, the integral in (\ref{latetimeG}) does not seem to be exactly doable. To derive an approximation, we write $w=\sin \theta, \gamma_1=\gamma_0 \sin \phi, \gamma=\gamma_0 \cos \phi$ with $0 <\phi \leq \pi/2$. Then, for $\gamma_0 t \gg 1$, the integrand can be approximated by a Gaussian:
\bea
\int_{\frac{a+|x|}{v_0 t}}^1 dw e^{-\gamma_0 t} e^{\gamma_1 t w} \frac{e^{t \sqrt{(\gamma_0^2-\gamma_1^2) (1-w^2)}}}{(1-w^2)^{1/4}} 
&\approx& \int_{-\Phi}^\infty d\theta \sqrt{\cos (\theta+\phi)} e^{-\frac{\gamma_0 t \theta^2}{2}} \\
&\approx& \sqrt{\frac{2 \pi \cos \phi}{\gamma_0 t}} -\frac{\sqrt{\cos(\phi-\Phi)}}{\gamma_0 t \Phi} e^{-\frac{\gamma_0 t \Phi^2}{2}} 
\eea
where, $\Phi \approx \phi-\frac{a+|x|}{v_0 t}$. At large times where $v_0 t \gg a+|x|$, using the above approximation in (\ref{latetimeG}), we arrive at an expression similar to (\ref{g0g1Gfn}) for $\gamma_0=\gamma_1$, 
\bea
G(x,t; a)   
&\approx& -\frac{v_0}{2} e^{\frac{\gamma_1}{v_0} (a-|x|)} \frac{e^{-(\gamma_0-\sqrt{\gamma_0^2-\gamma_1^2})t}}{\sqrt{2 \pi t \sqrt{\gamma_0^2-\gamma_1^2}}}  [\Theta(v_0 t-|x-a|)-\Theta(v_0 t-|x| -a)] \nn \\
&+& G(x)  \Theta(v_0 t-|x| -a) 
\eea
where, $G(x)$ is the stationary state Green's function (\ref{stepGss}). At larger times, the above expression suggests that the stationary state is approached over a time scale, $(\gamma_0 -\sqrt{\gamma_0^2-\gamma_1^2})^{-1}$ (with a $\ln t$ correction due to $t^{-1/2}$ factor). Although  the amplitude of the mean and variance calculated from the above expression does not match the simulations, the relaxation time scale is consistent with the numerical results in Fig.~\ref{fig:u0} and that obtained in \cite{Singh:2020}.

\clearpage

%


\end{document}